%% file: 0_Main.tex
\documentclass[preprint,pre,floats,aps,amsmath,amssymb]{revtex4}

\usepackage[percent]{overpic}

\usepackage{bm}
\usepackage[colorlinks=true,allcolors=blue]{hyperref}
\usepackage{graphicx}
\usepackage{cleveref}
\usepackage{mathtools} 
\usepackage{bm}
\usepackage{tabularx}
\usepackage{float}
\usepackage{lipsum} 
\usepackage{marginnote}
\usepackage{subcaption}
\usepackage{array}
\usepackage{booktabs}
\usepackage{multirow}

\usepackage{color, soul}
\usepackage[normalem]{ulem}
\usepackage{multirow}
\usepackage{amsfonts}
\newcommand{\overbar}[1]{\mkern 1.5mu\overline{\mkern-1.5mu#1\mkern-1.5mu}\mkern 1.5mu}
\setstcolor{red}

\newcommand\redsout{\bgroup\markoverwith{\textcolor{red}{\rule[0.5ex]{6pt}{1.5pt}}}\ULon}

\begin{document}                  

\title{Effect of negative triangularity on SOL plasma turbulence in double-null L-mode plasmas}
\author{K. Lim$^{1,2}$\footnote{Email: kyungtak.lim@ntu.edu.sg}, P. Ricci$^2$, L. Lebrun$^2$}
\affiliation{$^1$  School of Physical and Mathematical Sciences, Nanyang Technological University, 637371 Singapore, $^2$ École Polytechnique Fédérale de Lausanne (EPFL), Swiss Plasma Center (SPC), EPFL SB, Station 13, CH-1015 Lausanne, Switzerland}

\begin{abstract}
The effects of negative triangularity (NT) on boundary plasma turbulence in double-null (DN) configurations are investigated using global, nonlinear, three-dimensional, flux-driven two-fluid simulations. NT plasmas exhibit suppressed interchange-driven instabilities, resulting in enhanced confinement and lower fluctuation levels compared to positive triangularity (PT) plasmas. This reduction in interchange instability is associated with the weakening of curvature effects in the unfavorable region, caused by the stretching of magnetic field lines at the outer midplane. The magnetic disconnection between the turbulent low-field side (LFS) and the quiescent high-field side (HFS) results in most of the heat flux reaching the DN outer targets. In NT plasmas, the power load on the outer target is reduced, while it increases on the inner target, indicating a reduced in-out power asymmetry compared PT plasmas. Furthermore, the analysis of power load asymmetry between the upper and lower targets shows that the up-down power asymmetry is mitigated in NT plasmas, mainly due to the reduced total power crossing the separatrix. The reduction of interchange instabilities in NT plasmas also affects the blob dynamics. A three-dimensional blob analysis reveals that NT plasmas feature smaller blob sizes and slower propagation velocities. Finally, an analytical scaling law for blob size and velocity that includes plasma shaping effects is derived based on the two-region model and is found to qualitatively capture the trends observed in nonlinear simulations.
\end{abstract}

\maketitle

\input{1_Intro}

\input{2_Model}
\input{3_Simulation_setup}
\input{4_Lp_estimate}
\input{5_Power_load}

\input{6_Blobs}

\input{7_Conclusion}
\begin{acknowledgments}
This work has been carried out within the framework of the EUROfusion Consortium, via the Euratom Research and Training Programme (Grant Agreement No 101052200 — EUROfusion) and funded by the Swiss State Secretariat for Education, Research and Innovation (SERI). This research is also supported by the National Research Foundation, Singapore. Views and opinions expressed are however those of the author(s) only and do not necessarily reflect those of the European Union, the European Commission, or SERI. Neither the European Union nor the European Commission nor SERI can be held responsible for them. The simulations presented herein were carried out in part on the CINECA Marconi supercomputer under the TSVV-1 and TSVV-2 projects. We acknowledge EuroHPC Joint Undertaking for awarding the project ID EHPC-EXT-2023E01-019 access to LUMI at CSC, Finland.
\end{acknowledgments}
\newpage
\bibliographystyle{ieetr} 
\bibliographystyle{unsrt}
\bibliography{9_bib.bib}
\end{document}

%% file: 1_Intro.tex
\section{Introduction}\label{Sec:1}
In recent years, extensive research has been carried out to explore alternative divertor configurations (ADCs) to the single-null (SN) configuration considered for ITER, where high core plasma performance can be maintained while ensuring that heat flux levels at the targets remain within material constraints \cite{Soukhanovskii2013, Reimerdes2020, Militello2021}. Among various ADCs, the double-null (DN) configuration is particularly interesting \cite{Wenninger2018, Brunner2018}. For instance, the presence of two X-points enables power control through two radiation fronts. Furthermore, DN configurations facilitate easier access to H-mode \cite{Meyer2005} and allow for the safe installation of external heating systems on the quiescent high-field side (HFS) region, due to its magnetic disconnection from the turbulent low-field side (LFS) \cite{Smick2013, LaBombard2017}. 

DN configurations can be combined with operation in negative triangularity (NT) scenarios \cite{Kikuchi2015}, which have recently emerged as a promising alternative to the H-mode operation in positive triangularity (PT) \cite{Kikuchi2015}. Experimental observations from various tokamaks have shown that NT L-mode plasmas can achieve H-mode-like confinement due to the reduced turbulence \cite{Austin2019, Coda2022} with intrinsically ELM-free scenarios \cite{Austin2019, Marinoni2019, Coda2022, Happel2023}. Moreover, recent numerical studies, including gyrokinetic \cite{Balestri2024, Mariani2024, Marinoni2024, DiGiannatale2024} and fluid modeling \cite{Riva2017, Muscente2023, Lim2023, Tonello2024}, provide evidence of reduced plasma turbulence when NT is considered. These findings underscore the viability of NT plasmas as an effective solution for power handling in fusion devices compared to the PT H-mode plasmas.
 
In this work, we investigate the effects of triangularity on SOL plasma turbulence in DN L-mode plasmas. We first observe that SOL plasma turbulence is reduced in NT plasmas, due to the decreased interchange drive and the reduced area of the magnetic surface in the bad curvature region. Second, we compare the characteristic pressure length scale $L_p$ between NT and PT plasmas, disentangling the impact of triangularity on $L_p$. Third, a comparison of the up-down power sharing asymmetry between the predictive scaling law derived in Ref.~\cite{Lim2024} and numerical measurements demonstrate that NT-DN plasmas exhibit less up-down asymmetry. Finally, we perform a three-dimensional analysis of the blob dynamics using a blob detection and tracking technique, comparing the behavior of blobs in NT and PT plasmas. In particular, we derive a theoretical scaling law for blob size and velocity that includes the effects of plasma shaping, which is found to be qualitatively consistent with numerical observations. Ultimately, our analysis enables us to assess the potential advantages of NT plasmas in DN configuration for power handling.

In the present study, we leverage previous turbulent simulations in DN scenarios \cite{Lim2024} performed using the GBS code. These simulations identified the physical mechanisms determining the power-sharing asymmetry between upper and lower outer targets in both balanced and unbalanced DN configurations. We expand this analysis by examining the effects of triangularity on SOL plasma turbulence in DN L-mode plasmas, varying plasma resistivity, heating power, and triangularity.

The structure of the present paper is as follows. Section~\ref{Sec2} introduces the GBS model, while Section~\ref{Sec3} describes the numerical setup parameters and the magnetic equilibria used for NT and PT plasmas simulations in DN configurations. Section~\ref{Sec4} presents the numerical results, highlighting the reduction of SOL plasma turbulence and comparing analytical estimates for the pressure gradient length with nonlinear GBS simulations. Section~\ref{Sec5} discusses the inner-outer and upper-lower power asymmetry at the DN targets between NT and PT plasmas. Section~\ref{Sec6} focuses on the blob analysis, illustrating the differences between NT and PT plasmas, and derives an analytical scaling law for blob size and velocity that accounts for plasma shaping effects. Finally, Section~\ref{Sec7} summarizes the conclusions.

%% file: 2_Model.tex
\section{Numerical model}\label{Sec2}
The investigations presented in this work are based on the three-dimensional drift-reduced Braginskii equations \cite{Zeiler1997} implemented in the GBS code \cite{Ricci2012, Giacomin2021}. GBS uses a coordinate system independent of the magnetic field, allowing for the study of various magnetic configurations, such as the snowflake \cite{Giacomin2020_2}, negative triangularity \cite{Riva2017, Lim2023}, double-null \cite{Beadle2020, Lim2024}, and non-axisymmetric \cite{Coelho2022} configurations. It is therefore an ideal tool for investigating the effects of plasma shaping on plasma turbulence.

In the present study, we focus on an axisymmetric magnetic field in the electrostatic limit without including the interaction between plasma and neutrals, although these are implemented in GBS \cite{Giacomin2021}. As a result, the GBS model equations can be reformulated as follows:
\begin{widetext}
\begin{align}
     \frac{\partial n}{\partial t} = -\frac{\rho_*^{-1}}{B}[\phi, n] + \frac{2}{B}\bigg[C(p_e)-nC(\phi)\bigg]-\nabla_\parallel (nv_{\parallel e}) + D_n \nabla^2_\perp n + s_n,
\label{GBS_density}
\end{align}

\begin{align}
    \frac{\partial \Omega}{\partial t} = -\frac{\rho_*^{-1}}{B} \nabla \cdot [\phi, \omega] -\nabla \cdot \big(v_{\parallel i} \nabla_\parallel\omega\big) + B^2 \nabla_\parallel j_\parallel + 2B C(p_e + \tau p_i) + \frac{B}{3}C(G_i) + D_\Omega \nabla_\perp^2 \Omega,
\label{GBS_vorticity}
\end{align}
\begin{align}
    \frac{\partial v_{\parallel i}}{\partial t} = -\frac{\rho_*^{-1}}{B}[\phi, v_{\parallel i}]-v_{\parallel i} \nabla_\parallel v_{\parallel i} - \frac{1}{n}\nabla_\parallel (p_e + \tau p_i) -\frac{2}{3n}\nabla_\parallel G_i + D_{v_{\parallel i}}\nabla_\perp^2 v_{\parallel i},
\end{align}
\begin{align}
     \frac{\partial v_{\parallel e}}{\partial t} &= -\frac{\rho_*^{-1}}{B}[\phi, v_{\parallel, e}] - v_{\parallel e}\nabla_\parallel v_{\parallel e} + \frac{m_i}{m_e}\Bigg(\nu j_\parallel + \nabla_\parallel \phi -\frac{1}{n}\nabla_\parallel p_e-0.71 \nabla_\parallel T_e -\frac{2}{3n}\nabla_\parallel G_e \Bigg) \nonumber \\
    & + D_{v_{\parallel e}}\nabla_\perp^2 v_{\parallel e}, 
    \label{GBS_ve}
\end{align}
\begin{align}
     \frac{\partial T_i}{\partial t} &=-\frac{\rho_*^{-1}}{B}[\phi, T_i] - v_{\parallel i}\nabla_\parallel T_i + \frac{4}{3}\frac{T_i}{B}\Bigg[C(T_e) + \frac{T_e}{n}C(n)-C(\phi) \Bigg] -\frac{10}{3}\tau \frac{T_i}{B}C(T_i) \nonumber \\
     &+ \frac{2}{3}T_i\Bigg[(v_{\parallel i}-v_{\parallel e})\frac{\nabla_\parallel n}{n}-T_i \nabla_\parallel v_{\parallel e}\Bigg] + 2.61\nu n (T_e -\tau T_i)+ \nabla_\parallel( \chi_{\parallel i}\nabla_\parallel T_i) \nonumber\\
     &+ D_{T_i}\nabla_\perp^2 T_i + s_{T_i},
\end{align}
\begin{align}
     \frac{\partial T_e}{\partial t} &= -\frac{\rho_*^{-1}}{B}[\phi, T_e] - v_{\parallel e}\nabla_\parallel T_e + \frac{2}{3}T_e\Bigg[0.71 \frac{\nabla_\parallel j_\parallel}{n}-\nabla_\parallel v_{\parallel e}\Bigg] -2.61\nu n (T_e-\tau T_i)
    \nonumber \\
    &+ \frac{4}{3}\frac{T_e}{B}\Bigg[\frac{7}{2}C(T_e)+\frac{T_e}{n}C(n)-C(\phi) \Bigg] + \nabla_\parallel (\chi_{\parallel e}\nabla_\parallel T_e) + D_{T_e}\nabla_\perp^2 T_e + s_{T_e}. \label{GBS_electron_temperature}
\end{align}
\end{widetext}
The above equations are coupled with the Poisson equation, which avoids the Boussinesq approximation.
\begin{align}
    \nabla \cdot \bigg( n\nabla_\perp \phi \bigg) = \Omega - \tau \nabla_\perp^2 p_i,
\label{GBS_Poisson}
\end{align}
where $\Omega=\nabla \cdot \omega = \nabla \cdot (n\nabla_\perp \phi + \tau \nabla_\perp p_i)$ is the scalar vorticity.

In Eqs.~(\ref{GBS_density}--\ref{GBS_Poisson}), the plasma variables are normalized with respect to the reference values. Specifically, the plasma density, $n$, the ion and electron temperatures, $T_i$ and $T_e$, the ion and electron parallel velocities, $v_{\parallel i}$ and $v_{\parallel e}$, and the electric potential, $\phi$, are normalized to $n_0$, $T_{i0}$, $T_{e0}$, $c_{s0}=\sqrt{T_{e0}/m_i}$, and $T_{e0}/e$, respectively. Perpendicular lengths are normalized to the ion sound Larmor radius $\rho_{s0}=c_{s0}/\Omega_{ci}$, where $\Omega_{ci}=eB/m_i$ is the ion cyclotron frequency. Parallel lengths are normalized to the tokamak major radius, $R_0$. Time is normalized to $R_0/c_{s0}$.

The dimensionless parameters that govern the plasma dynamics in Eqs.~(\ref{GBS_density}--\ref{GBS_Poisson}) include the normalized ion Larmor radius $\rho_*=\rho_{s0}/R_0$, the ratio of ion to electron temperature $\tau=T_{i0}/T_{e0}$, and the normalized ion and electron viscosities $\eta_{0i}=0.96 n T_i \tau_i$ and $\eta_{0e}=0.73 n T_e \tau_e$, respectively, where $\tau_{e,i}$ represent the electron and ion collision times. The normalized ion and electron parallel thermal conductivities are defined as:
\begin{align}
    \chi_{\parallel i} = \bigg(\frac{1.94}{\sqrt{2\pi}}\sqrt{m_i} \frac{(4\pi \epsilon_0)^2}{e^4} \frac{c_{s0}}{R_0}\frac{T_{e0}^{3/2}\tau^{5/2}}{\lambda n_0} \bigg) T_i^{5/2}
\end{align}
and 
\begin{align}
    \chi_{\parallel e} = \bigg(\frac{1.58}{\sqrt{2\pi}}\frac{m_i}{\sqrt{m_e}}\frac{(4\pi \epsilon_0)^2}{e^4} \frac{c_{s0}}{R_0}\frac{T_{e0}^{3/2}}{\lambda n_0} \bigg) T_e^{5/2}.
\end{align}
The normalized Spitzer resistivity is defined as $\nu=e^2n_0R_0/(m_ic_{s0}\sigma_\parallel)=\nu_0 T_e^{-3/2}$, with
\begin{align}
    \sigma_\parallel = \Bigg(1.96\frac{n_0 e^2 \tau_e}{m_e}\Bigg)n = \Bigg(\frac{5.88}{4\sqrt{2\pi}} \frac{(4\pi \epsilon_0)^2}{e^2}\frac{T_{e0}^{3/2}}{\lambda \sqrt{m_e}}\Bigg)T_e^{3/2}
\end{align}
and
\begin{align}
    \nu_0 =\frac{4\sqrt{2\pi}}{5.88} \frac{e^4}{(4\pi \epsilon_0)^2}\frac{\sqrt{m_e}R_0 n_0 \lambda}{m_i c_{s0}T_{e0}^{3/2}},
\label{Collisionality}
\end{align}
where $\lambda$ is the Coulomb logarithm.

The differential geometrical operators used in Eqs.~(\ref{GBS_density}--\ref{GBS_Poisson}) are defined as follows:
\begin{align}
    [\phi,f] &= \bm{b} \cdot (\nabla \phi \times \nabla f), \label{GBS_operator1}\\
    \mathcal{C}(f) &= \frac{B}{2}\bigg( \nabla \times \frac{\bm{b}}{B}\bigg) \cdot \nabla f, \label{GBS_operator2} \\
    \nabla_\parallel f &= \bm{b} \cdot \nabla f, \\
    \nabla_\perp^2 f &=\nabla \cdot \big[(\bm{b}\times \nabla f)\times \bm{b}\big].\label{GBS_operator4}
\end{align}
These correspond to the $E \times B$ convective term, the curvature operator, the parallel gradient, and the perpendicular Laplacian of a scalar function $f$. To compute these differential operators, the non-field-aligned cylindrical coordinates $(R, \varphi, Z)$ are used, where $R$ represents the radial distance from the torus axis of symmetry, $Z$ the vertical coordinate, and $\varphi$ the toroidal angle. An axisymmetric magnetic field is considered, expressed as $\bm{B}=RB_\varphi\nabla \varphi + \nabla \varphi \times \nabla \psi$, where $\psi(R,Z)$ is the poloidal magnetic flux. Note that these operators depend on the sign of the normalized magnetic field $\bm{b}=\bm{B}/B$, which determines whether the magnetic drifts are directed toward the lower X-point (favorable direction) or away from it (unfavorable).

The gyroviscous terms are defined as 
\begin{eqnarray}
    G_i=-\eta_{0i}\Bigg[2\nabla_\parallel v_{\parallel i} + \frac{1}{B}C(\phi) + \frac{1}{enB}C(p_i) \Bigg]
\end{eqnarray}
and
\begin{eqnarray}
    G_e=-\eta_{0e}\Bigg[2\nabla_\parallel v_{\parallel e} + \frac{1}{B}C(\phi)-\frac{1}{enB}C(p_e) \Bigg].
\end{eqnarray}
To improve the numerical stability of the simulation, artificial diffusion terms $D_f\nabla_\perp^2 f$ are introduced on the right-hand side of Eqs.~(\ref{GBS_density}--\ref{GBS_Poisson}).

To mimic the ionization process and Ohmic heating within the core, toroidally uniform density and temperature sources are applied inside the last closed flux surface (LCFS). These sources are expressed as follows:
\begin{align}
    s_n &= s_{n0}\exp{\Bigg( -\frac{[\psi(R,Z)-\psi_n]^2}{\Delta_n^2}\Bigg)},\label{density_source} \\
    s_T &= \frac{s_{T0}}{2}\Bigg[\tanh\Bigg(-\frac{\psi(R,Z)-\psi_T}{\Delta_T}\Bigg)+1\Bigg], \label{temperature_source}
\end{align}
where $\psi_n$ and  $\psi_T$ represent the flux function values inside the LCFS, determining the radial position of the sources terms, while $\Delta_n$ and $\Delta_T$ set their radial width. 

Boundary conditions that satisfy the Bohm-Chodura criterion are implemented at the magnetic pre-sheath entrance \cite{Loizu2012}. By neglecting the gradients of density and electrostatic potential in directions tangent to the wall, these boundary conditions take the following form:
\begin{align}
    v_{\parallel i}&=\pm \sqrt{T_e+\tau T_i}, \\
    v_{\parallel e}&=\pm\sqrt{T_e + \tau T_i}\exp{\Bigg( \Lambda - \frac{\phi}{T_e}\Bigg)}, \\
    \partial_Z n &= \mp \frac{n}{\sqrt{T_e+\tau T_i}}\partial_Z v_{\parallel i}, \\
    \partial_Z \phi &= \mp \frac{T_e}{\sqrt{T_e + \tau T_i}}\partial_Z v_{\parallel i},\\
    \partial_Z T_e &= \partial_Z T_i = 0, \\
    \omega &= -\frac{T_e}{T_e + \tau T_i}\bigg[(\partial_Z v_{\parallel i})^2 \pm \sqrt{T_e + \tau T_i} \partial^2_Z v_{\parallel i}\bigg].
\end{align}
In these equations, the $\pm$ sign indicates whether the magnetic field line enters (top sign) or leaves (bottom sign) the wall, and $\Lambda = \log \sqrt{m_i/(2\pi m_e)}\simeq 3$. For the left and right domain boundaries, the electric potential is defined as $\phi=\Lambda T_e$, and the derivatives normal to the wall are set to vanish for all other quantities.

%% file: 3_Simulation_setup.tex
\section{Simulation setup}\label{Sec3}
\begin{figure}[H]
\begin{center}
\includegraphics[width=\textwidth]{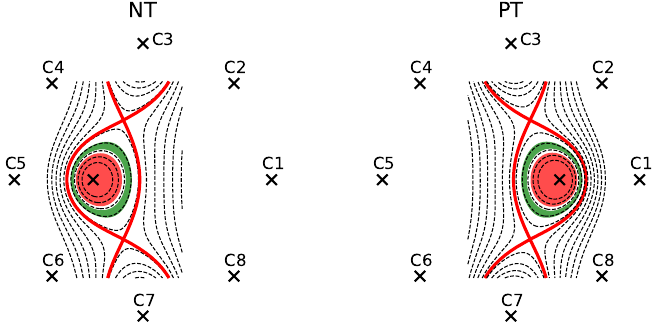}
\caption{Examples of magnetic equilibria of NT and PT plasmas in DN configurations used for the nonlinear GBS simulations. The black crosses mark the positions of the current-carrying coils (C1-C8) used to generate different values of $\delta$. The red solid line represents the separatrix. The red shaded region indicates the area where the heating source is applied and, similarly, the green shaded region represents the density source, mimicking the ionization processes occurring inside the separatrix}
\label{Fig:Equil_DN_tria}
\end{center}
\end{figure}

The poloidal magnetic flux, $\psi(R,Z)$, is evaluated assuming a Gaussian-like current centered at the magnetic axis, along with additional current-carrying wires (C1-C8) positioned outside the simulation box. While the position of these coils remains fixed throughout this study, the coil currents are adjusted to achieve different values of triangularity, while maintaining a constant elongation, as shown in Figure \ref{Fig:Equil_DN_tria}. 

For all GBS simulations presented in this study, we adopt a parameter setup for DN configurations similar to that described in Ref.~\cite{Lim2024}. Specifically, we consider a numerical grid defined by $(N_R, N_Z, N_\varphi)=(240,320, 80)$ for a simulation with domain size $(L_R, L_Z)=(600\rho_{s0}, 800\rho_{s0})$ and a time step of $\Delta t=10^{-5}R_0/c_{s0}$. We use $\rho_*=\rho_{s0}/R_0=1/700$, corresponding to approximately one-third the size of TCV tokamak \cite{Reimerdes2022}. The radial width and position of the source terms are set to ($\Delta_n, \Delta_T) = (500, 700)$ and $(\psi_n, \psi_T)=(140, 160)$ in Eqs.~(\ref{density_source}--\ref{temperature_source}), as illustrated in the shaded region of Figure \ref{Fig:Equil_DN_tria}. The heating source amplitude, $s_{T0}$, is applied equally to both ion and electron species.

To explore the effects of triangularity on plasma turbulence in DN configurations, we carry out a parametric scan by varying triangularity $\delta = \{\pm 0.3, \pm 0.4, \pm 0.5\}$, plasma resistivity $\nu_0=\{0.1, 0.3, 1.0\}$, and the amplitude of heating power $s_{T0}=\{ 0.05, 0.15, 0.3\}$ across both NT and PT plasmas, while maintaining a constant elongation $\kappa=1.6$. We note that, according to Eq.~(\ref{Collisionality}), a variation of the plasma resistivity is equivalent to a variation of the plasma density. To reduce the computational cost of these simulations and allow for an extensive parametric scan, we use a mass ratio of $m_i/m_e=200$. This reduced ion-to-electron mass ratio does not significantly affect the results of nonlinear simulations in L-mode plasmas scenarios, as plasma resistivity dominates over inertial effects when $\nu_0>(m_e/m_i)\gamma$, with $\gamma$ denoting the growth rate. To facilitate the analysis, we keep the plasma parameters $\tau=T_{i0}/T_{e0}=1$, $\eta_{0e}=\eta_{0i}=1$ and $ \chi_{\parallel e}=\chi_{\parallel i}=1$, constant throughout our study. The safety factor is adjusted to $q_0 \simeq 1$ at the magnetic axis and $q_{95}\simeq 4$ at the tokamak edge in both NT and PT plasmas. Additionally, the reference toroidal magnetic field $B_T$ is set for all simulations such that the ion-$\nabla B$ drift is away from the lower X-point. 

We use the simulation with parameters $\nu_0=0.3, s_{T0}=0.15$, and $\delta=\pm 0.5$ as the reference case. Within the selected ranges for plasma resistivity and heating source amplitude, plasma turbulence is driven by resistive ballooning modes (RBMs). These modes are destabilized by the magnetic field curvature and the plasma pressure gradient \cite{Zeiler1997}. The dominant RBMs in the SOL region are equivalent to those observed in tokamak L-mode operation \cite{Fundamenski2007, Mosetto2013, Giacomin2020}, and expected in NT plasma scenarios \cite{Kikuchi2019}. 

The simulations are carried out until they reach a quasi-steady state, where the input power is balanced by perpendicular transport and losses at the vessel wall. The analysis of the simulations is carried out once this quasi-steady state is achieved, by averaging the plasma quantities over a time window of $10t_0$. In this paper, we denote time- and toroidally- averaged values with an overline and fluctuations with a tilde, such that ${f}=\bar{{f}}+\tilde{{f}}$ for a generic quantity ${f}$.

%% file: 4_Lp_estimate.tex
\section{Estimate of the pressure gradient length}\label{Sec4}

The pressure gradient length in the near SOL, $L_{p}=-p_e/\nabla p_e$, is determined by the level of SOL plasma cross-field transport and correlates with the power fall-off decay length $\lambda_q$ at the outer targets \cite{Giacomin2021_2, Lim2023}. The theoretical scaling law for estimating $L_p$, which includes shaping parameters in the SN configuration, was derived in Ref.~\cite{Lim2023} and validated against DN configurations, demonstrating its applicability to both configurations \cite{Lim2024}. NT configurations exhibit a steeper plasma pressure profile than PT configurations due to a lower level of turbulence and associated transport. 

In this section, we compare the electron pressure, $p_e$, obtained from nonlinear simulations, between NT and PT plasmas. In addition, we verify that SOL turbulence in the L-mode plasmas considered in the present study is mainly driven by RBMs, which is a key assumption underlying the analytical derivation of $L_p$. Finally, we compare the $L_p$ values derived from the analytical expression with those obtained from nonlinear simulations.

\begin{figure}[H]
\begin{center}
\includegraphics[width=\textwidth]{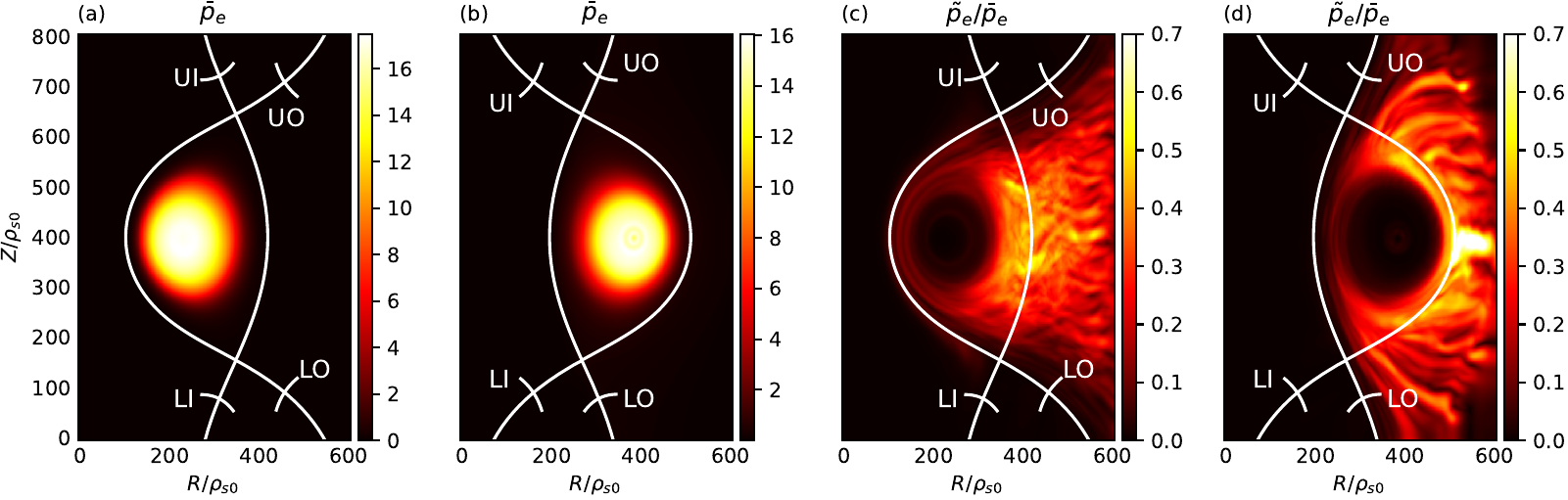}
\caption{Equilibrium electron pressure, $p_e$, for (a) the NT and (b) PT configurations, and snapshots of turbulent fluctuations for (c) the NT and (d) PT configurations. The reference simulations with $\nu_0=0.3$, $s_{T0}=0.15$, using $\delta=-0.5$ for (a) and (c) and $\delta=+0.5$ for (b) and (d) are considered. Four solid lines are positioned in front of each divertor target, labeled as Lower Outer (LO), Upper Outer (UO), Lower Inner (LI), and Upper Inner (UI), where the target heat flux is evaluated.}
\label{Fig:Snapshot}
\end{center}
\end{figure}

Figure~\ref{Fig:Snapshot} illustrates equilibrium and typical snapshots of the fluctuations of the electron pressure, $p_e$, in NT and PT plasmas. Similar to the SN configurations \cite{Lim2023}, NT plasmas exhibit a relatively higher equilibrium pressure, $\overbar{p}_e$, and reduced fluctuation levels, $\widetilde{p}_e$, in the SOL region compared to PT plasmas. The fluctuating components show that plasma turbulence develops inside the separatrix and propagates from the near to the far SOL regions due to the presence of blobs, eventually reaching the right wall of the simulation domain. The presence of a secondary X-point in DN configurations results in a quiescent plasma at the HFS region, which is magnetically disconnected from the turbulent LFS region.

The lower fluctuation levels in NT plasmas are associated with a reduction of interchange-driven instabilities within the bad curvature region \cite{Lim2023}. This reduction occurs mainly because particle trajectories remain longer in the bad curvature region of PT plasmas, making RBMs more susceptible to destabilization in this region \cite{Riva2017}. In addition, the poloidal length of the separatrix in the bad curvature region is shorter in NT plasmas ($\sim500\rho_{s0}$) compared to PT cases ($\sim700\rho_{s0}$), allowing more surface RBMs to develop.

\begin{figure}[H]
\begin{center}
\includegraphics[width=\textwidth]{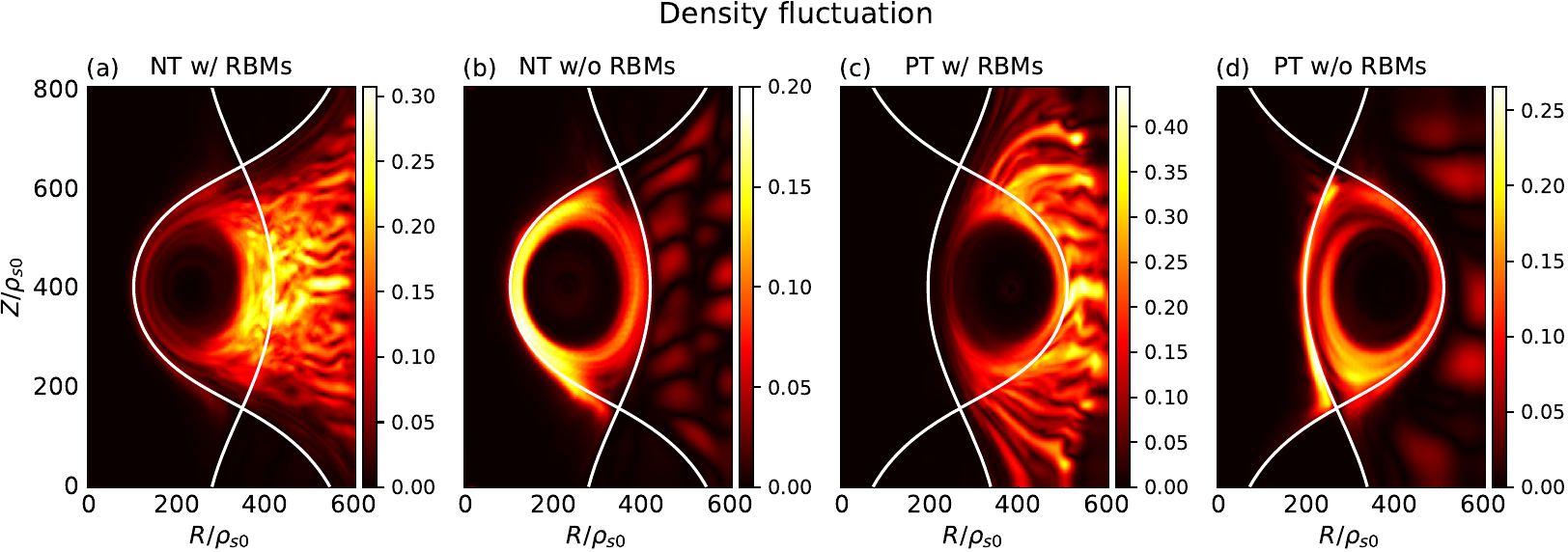}
\caption{2D snapshots of density fluctuations from the reference simulations: (a) NT with RBMs drive, (b) NT without RBMs drive, (c) PT with RBMs drive, (d) PT without RBMs drive. The curvature operator in the vorticity equation from Eq.~(\ref{GBS_vorticity}) is zeroed out remove RBMs drive.}
\label{Fig:DN_fluct_no_curv}
\end{center}
\end{figure}

To verify the hypothesis that RBMs dominate our L-mode plasmas and that the suppressed turbulence in NT plasmas is linked to a reduction in interchange instabilities, we zero out the curvature operator in the vorticity equation, Eq.~(\ref{GBS_vorticity}), which constitutes the RBM drive, and compare its impact on SOL plasma turbulence. Figure~\ref{Fig:DN_fluct_no_curv} presents 2D snapshots of density fluctuations in NT and PT plasmas with and without the RBM drive. First, consistent with Figure~\ref{Fig:Snapshot}, PT plasmas display higher levels of density fluctuations compared to NT plasmas (Figure~\ref{Fig:DN_fluct_no_curv}\textcolor{blue}{a} and \ref{Fig:DN_fluct_no_curv}\textcolor{blue}{c}). Second, we observe a significant disappearance of the fluctuating structures in both the near and far SOL regions where the RBM drive is removed (Figure~\ref{Fig:DN_fluct_no_curv}\textcolor{blue}{b} and \ref{Fig:DN_fluct_no_curv}\textcolor{blue}{d}). This suggests that the simulations carried out in the L-mode plasma conditions considered here are dominated by RBMs, which is the primary assumption behind the analytical derivations that follow, particularly for the $L_p$ gradient length.

The analytical derivation of $L_p$ is based on a gradient removal theory \cite{Ricci2008, Ricci2013}, where the local flattening of the plasma pressure profile provides the main mechanism for the saturation of the growth of linear instabilities that drive turbulence. The value of $L_p$ is then determined by balancing perpendicular turbulent transport with parallel losses at the ends of the magnetic field lines. A detailed derivation of $L_p$ can be found in Ref.~\cite{Lim2023}. The analytical expression for the $L_p$ estimate in DN configurations, which accounts for elongation and triangularity effects, is given by:

\begin{align}
    L_{p} \sim \mathcal{C}(\kappa, \delta, q) \bigg[ \rho_* (\nu_0 \bar{n} q^2)^2 \bigg(\frac{L_\chi \bar{p}_e}{S_p}\bigg)^{4} \bigg]^{1/3},
\label{Lp_analytical}
\end{align}
where $q$ is the safety factor at the tokamak edge, $L_\chi\simeq \pi a (0.45 + 0.55\kappa) + 1.33a\delta$ is an approximation of the poloidal length of the separatrix, and $S_p$ represents the volume-integrated power source within the separatrix \cite{Lim2023}. The curvature operator, $\mathcal{C}(\kappa, \delta, q)$, defined at the outer midplane $(\theta=0)$ where RBMs are mostly destabilized, is expressed as:
\begin{align}
    \mathcal{C}(\kappa, \delta, q) = 1-\frac{\kappa-1}{\kappa+1}\frac{3q}{q+2} + \frac{\delta q}{1+q} + \frac{(\kappa-1)^2(5q-2)}{2(\kappa+1)^2(q+2)} + \frac{\delta^2}{16}\frac{7q-1}{1+q}.
    \label{Curvature_operator}
\end{align}

The analytical $L_p$ scaling law in Eq.~(\ref{Lp_analytical}) has been validated against experimental multi-machine datasets, focusing also on plasma shaping effects within SN configurations \cite{Lim2023}, demonstrating its reliability in predicting $\lambda_q$ across both NT and PT plasmas. The $L_p$ estimate was applied to DN configurations \cite{Lim2024} and will be used to estimate $\lambda_q$ for both upper and lower outer targets. 

\begin{figure}[H]
\begin{center}
\includegraphics[width=\textwidth]{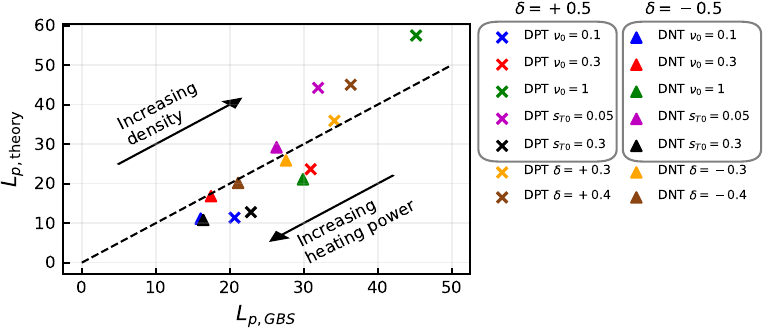}
\caption{Comparison of the pressure gradient length, $L_p$, between the analytical scaling law in Eq.~(\ref{Lp_analytical}) and the nonlinear GBS simulations. A scan of plasma resistivity $\nu_0$, heating power $s_{T0}$, and triangularity $\delta$ is carried out for both NT and PT plasmas. The $R^2$-score of the comparison is, approximately, $0.728$.}
\label{Fig:Lp_estimate}
\end{center}
\end{figure}

Figure~\ref{Fig:Lp_estimate} compares the analytical $L_p$ estimate derived in Eq.~(\ref{Lp_analytical}) with the $L_p$ values obtained from nonlinear GBS simulations. The $L_p$ values are calculated by considering the value of $p_e$ at the separatrix and evaluating its gradient over a radial interval extending $ 40\rho_{s0}$, centered at the separatrix. The analytical scaling law captures the $L_p$ trend with an $R^2$-score of 0.728. According to the analytical expression in Eq.~(\ref{Lp_analytical}), $L_p$ increases with plasma resistivity $\nu_0$ and decreases with input heating power $s_{T0}$. This relationship aligns with previous findings that higher resistivity and lower heating power enhance turbulence fluctuations and cross-field transport \cite{Giacomin2020}, which flattens the pressure profile in the near SOL and results in higher $L_p$ values. As shown in Figure~\ref{Fig:Snapshot}, reduced fluctuation levels in NT result in smaller $L_p$ and, overall, $L_p$ increases with $\delta$. 

%% file: 5_Power_load.tex
\section{Target heat load and in-out and up-down asymmetries}\label{Sec5}

Experimental observations from various tokamaks report a pronounced in-out \cite{Liu2014, Liu2016, Du2017}, as well as up-down power load asymmetry \cite{Morel1999, Petrie2001, Temmerman2011, Brunner2018, Fevrier2021}. Several key mechanisms have been suggested to explain these power-sharing asymmetries, including cross-field drifts \cite{Rognlien1999, Cohen1999, Rensink2000, Rubino2020}, ballooning modes \cite{Du2015}, Pfirsch-Schl\"uter (PS) flows \cite{Schaffer1997, Asakura2004}, flux compression between the two separatrices \cite{Osawa2023}, and different recycling rates at the divertor targets \cite{Rensink2000}.

Recent work with DN configurations using GBS suggests that the interplay between poloidal diamagnetic drift and radial turbulent transport is a key factor in determining the up-down power asymmetry, along with the magnetic imbalance \cite{Lim2024}. More precisely, poloidal diamagnetic drift tends to increase the up-down asymmetry, while radial turbulent transport reduces it. NT plasmas with DN configurations operating in L-mode scenarios \cite{Kikuchi2019} exhibit stronger plasma turbulence than H-mode discharges, which might lead to a lower level of up-down power load asymmetry, a beneficial effect for the operation of a fusion power plant.

We first estimate the effects of triangularity on the power load reaching the inner and outer target plates by considering the lower outer (LO) and lower inner (LI) targets shown in Figure~\ref{Fig:Snapshot}. These plates are located midway between the X-point and the wall to avoid numerical artifacts near the wall regions. The heat flux at the upper outer (UO) and upper inner (UI) targets are evaluated similarly.

\begin{figure}[H]
\begin{center}
\includegraphics[width=0.9\textwidth]{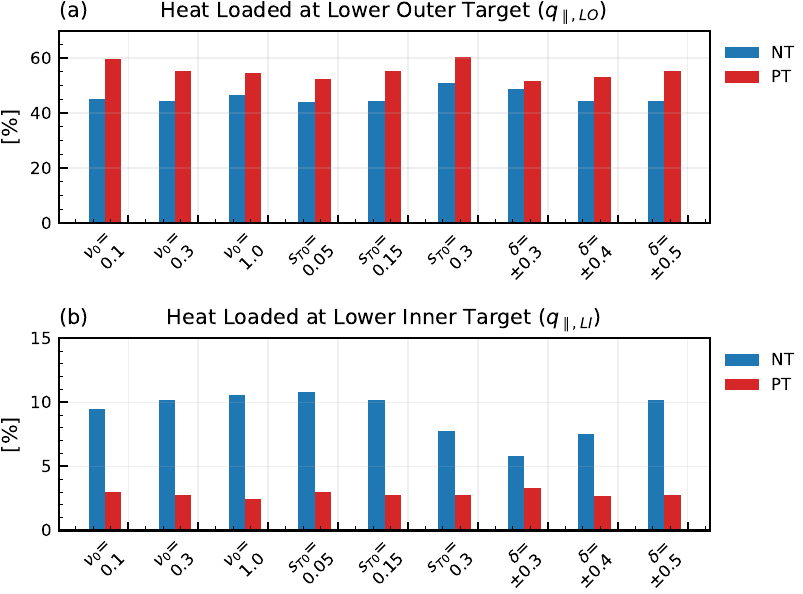}
\caption{Percentage of the heat flux at the (a) LO and (b) LI targets for NT (blue) and PT (red) plasmas. The parallel heat flux $q_\parallel$ is time-averaged over a time interval of 10$t_0$. A scan of plasma resistivity, heating power, and triangularities is considered.}
\label{Fig:qpar_lower}
\end{center}
\end{figure}

Figure~\ref{Fig:qpar_lower} presents the time-integrated parallel heat flux reaching the LO and LI targets, for simulations carried out with different values of plasma resistivity $\nu_0$, input heating power $s_{T0}$, and triangularity $\delta$. Consistent with previous results \cite{Lim2024}, the magnetic disconnection between the quiescent HFS and the turbulent LFS results in less than $90\%$ of the heat flux reaching the outer targets in both NT and PT plasmas. More precisely, across all values of $\nu_0$ and $s_{T0}$, PT plasmas (red) exhibit a larger heat flux at the outer target compared to NT plasmas (blue). Indeed, PT plasmas are characterized by a negligible inner heat flux, less than 5\%, whereas NT plasmas present a more significant inner heat flux, up to 10\%. Consequently, the in-out power asymmetry decreases in NT plasmas, while it becomes more pronounced in PT plasmas. 

The non-negligible power load at the inner targets in NT plasmas can be attributed to both geometrical factors and cross-field turbulent transport. First, as illustrated in Figure~\ref{Fig:Snapshot}, the longer separatrix at the HFS in NT plasmas leads to an increased heat flux crossing the HFS separatrix. Second, the reduced levels of cross-field turbulent transport at the LFS in NT plasmas contribute to a relatively significant power load at the inner targets, thereby reducing in-out power asymmetry in NT plasmas. Indeed, previous studies on the EAST tokamak and BOUT++ simulations \cite{Liu2012, Liu2014} have demonstrated that the in-out power asymmetry is proportional to the power crossing the separatrix, $P_{\textrm{SOL}}$. This suggests that enhanced SOL turbulence leads to a stronger in-out power asymmetry, consistent with the behavior observed in the PT plasmas considered in this study.

We now turn our attention to the up-down power sharing asymmetry. We apply the predictive analytical scaling law derived in Ref.~\cite{Lim2024} to evaluate the up-down asymmetry at the outer targets in DN configurations, exploring the effects of plasma triangularity. In balanced DN configurations with an inter-separatrix distance of $\delta R=0$ considered here, the scaling law proposed in Ref.~\cite{Lim2024} reduces to:
\begin{align}
      \lvert q_{\parallel, \textrm{LO}} - q_{\parallel, \textrm{UO}}\rvert
      =q_{\textrm{asym}} &= q_\psi \alpha_d K, 
      \label{scaling_heat_asymmetry2}
\end{align}
with
\begin{align}
    \alpha_d = \rho_*^{3/4} \bar{n}^{-3/2} \bar{p}_e L_p^{-1/4} \nu_0^{-1/2}q^{-1},
    \label{diag_analytical}
\end{align}
where $q_\psi$ is the heat flux crossing the separatrix in the LFS region, $\alpha_d$ is a dimensionless diamagnetic parameter that includes the effects of both diamagnetic drift and turbulence, $K$ is a numerical coefficient used to account for the order of magnitude estimates considered and determined by fitting simulations and experimental results. 

The parameter $\alpha_d$ in Eq.~(\ref{scaling_heat_asymmetry2}) represents the competing effects of the diamagnetic drift and turbulence in generating up-down heat asymmetry. Given the dominant RBMs nature of turbulence in our simulations, we define $\alpha_d=\lvert \bm{v}_d \rvert k_{\textrm{RBM}}/\gamma_{\textrm{RBM}}$ with $\bm{v}_d$ the diamagnetic velocity and the characteristic wavenumber and the growth rate of the RBMs, being defined as $k_{\textrm{RBM}}=1/\sqrt{\bar{n}\nu q^2 \gamma_{\textrm{RBM}}}$ and $\gamma_{\textrm{RBM}}=\sqrt{(2\bar{T}_e)/(\rho_* L_p)}$, respectively. We approximate $\lvert \bm{v}_d \rvert \simeq \bar{p}_e/(\bar{n} L_p)$. Higher $\alpha_d$ values indicate more pronounced asymmetry-driving mechanisms.

\begin{figure}[H]
\begin{center}
\includegraphics[width=\textwidth]{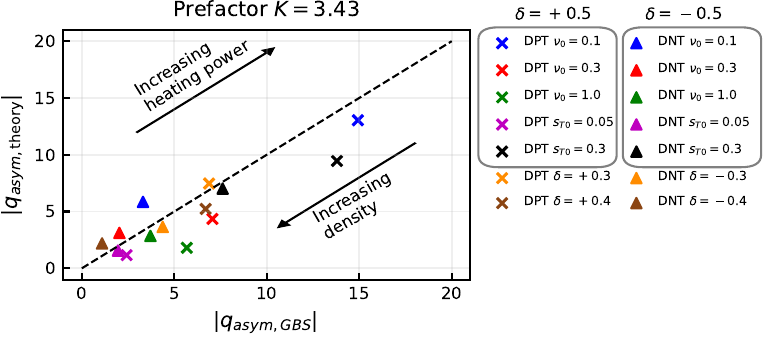}
\caption{Comparison of the heat flux asymmetry between the analytical scaling law in Eq.~(\ref{scaling_heat_asymmetry2}) and the nonlinear GBS simulations. We set $K=3.43$ for all the simulations, and obtain an $R^2$-score of 0.74.}
\label{Fig:Asymmetry_scaling}
\end{center}
\end{figure}

Figure~\ref{Fig:Asymmetry_scaling} compares the heat flux asymmetry, $q_{\rm{asym}}$, predicted by the analytical scaling law in Eq.~(\ref{scaling_heat_asymmetry2}) with the results from nonlinear GBS simulations. We set the prefactor $K=3.43$, determined using linear regression methods, whereas in Ref.~\cite{Lim2024}, the prefactor was set to $K = 2.84$. The difference is due to the variation in the parameter spaces explored in these works. Overall, the scaling law effectively captures the trend of heat asymmetry observed across various simulations. Consistent with Ref.~\cite{Lim2024}, increasing $\nu_0$ and decreasing $s_{T0}$ leads to a reduction of the up-down heat asymmetry, as a consequence of the decrease in $\alpha_d$ of Eq.~(\ref{scaling_heat_asymmetry2}). 

The clear difference in heat asymmetry between NT and PT plasmas is highlighted in Figure~\ref{Fig:Asymmetry_scaling}, with PT plasmas showing more pronounced heat asymmetry. This behavior is described by Eq.~(\ref{scaling_heat_asymmetry2}), where 
$q_{\textrm{asym}}$ is directly proportional to the heat flux crossing the separatrix in the LFS region, $q_\psi$, as well as the dimensionless diamagnetic parameter, $\alpha_d$. The reduced up-down heat asymmetry at the outer targets in NT plasmas can be attributed to two main factors. First, the scaling law in Eq.~(\ref{scaling_heat_asymmetry2}) does not account for inner heat loads. However, NT plasmas exhibit a non-negligible amount of inner heat load, which decreases the up-down asymmetry at the outer targets. Second, the heat flux crossing the separatrix in the LFS region is smaller in NT plasmas compared to PT plasmas, further reducing the observed up-down asymmetry. Consequently, although the smaller $L_p$ in NT plasmas leads to a larger $\alpha_d$, the combined effects of the inner heat load and reduced $q_psi$ eventually result in the mitigated up-down heat asymmetry at the outer targets observed in Figure~\ref{Fig:Asymmetry_scaling}.

It is important to note that the up-down asymmetry, $q_{\textrm{asym}}$, depends on the magnetic imbalance. Achieving an ideal equal distribution between upper and lower outer targets in future fusion devices may therefore require proper control of the DN magnetic imbalance.

%% file: 6_Blobs.tex
\section{Blob dynamics}\label{Sec6}
Blobs are coherent plasma structures that originate in the edge region, and move toward the far SOL \cite{Myra2006_2, Zweben2007, Ippolito2011}. Experimental observations clearly show that blobs typically form at the outer midplane (OMP) in the LFS, where the curvature-driven interchange instabilities are present \cite{Rudakov2002, Zweben2007}. This suggests a link between RBMs and blob formation, as well as the effect of triangularity on blob dynamics.

To investigate the blob dynamics \cite{Angus2012, Walkden2013, Easy2014}, three-dimensional simulations have been carried out using various codes, such as BOUT++ \cite{Russell2004}, GBS \cite{Nespoli2017, Paruta2019}, and GRILLIX \cite{Ross2019, Zholobenko2023}. Despite the extensive studies, the effect of triangularity on blob dynamics has yet to be investigated.

In this section, we present a three-dimensional blob analysis from nonlinear GBS simulations for PT and NT plasmas. We begin by applying blob detection techniques to identify blobs, as well as their sizes and velocities. The detected blobs are then compared with the two-region model \cite{Myra2006} to elucidate the main mechanisms behind the blob dynamics. Finally, we derive a new analytical scaling law for blob size and radial blob velocity by including the effects of triangularity, explaining the observed trends from the nonlinear GBS simulations.

\input{6-1_Blob_detection}

\input{6-2_Two_region}
\input{6-3_Scaling_law}

%% file: 6-1_Blob_detection.tex
\subsection{Blob detection and tracking}
To detect and track the motion of blobs in the GBS simulation, we use an image processing algorithm \cite{scikit-image}. We detect blobs as structures that meet the condition $n_{\rm{blobs}}(R, Z, t) > \overbar{n}(R, Z) + 2.5 \sigma_n(R, Z)$, where $\sigma_n$ is the standard deviation of the density. Consistent with previous GBS works in Refs.~\cite{Nespoli2017, Paruta2019}, once detected, a blob is tracked from one time frame to the next. However, in contrast to Refs.~\cite{Nespoli2017, Paruta2019} and for the sake of simplicity, we do not account for merging and splitting events. 

\begin{figure}[H]
    \centering
    \includegraphics[width=0.8\textwidth]{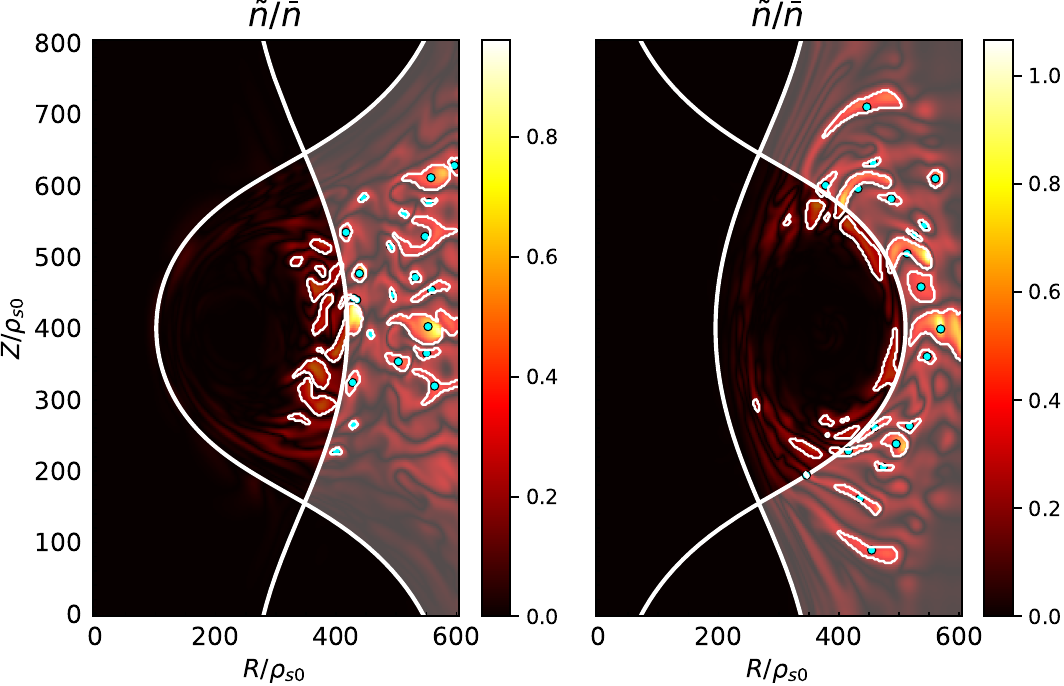}
    \caption{Two-dimensional snapshot of blob detection in NT and PT plasmas. Blobs that meet the threshold condition in the LFS region outside the separatrix (grey region) are detected, with their contours outlined as solid white lines. The center of mass of each detected blob is marked with a cyan dot.}
    \label{Fig:blob_detection}
\end{figure}

We focus our analysis on blob size and velocity, highlighting the differences between NT and PT plasmas in DN configurations, considering only blobs located outside the separatrix, as shown by the grey-shaded region in Figure~\ref{Fig:blob_detection}. Blobs are approximated as circles, centered at the blob center of mass location. The blob velocity is then determined by evaluating the distance traveled by the center of mass during the blob analysis time window, $t_{\rm{blobs}}$.

Figure~\ref{Fig:blob_detection} shows the result of the blob detection technique on a typical snapshot of the density fluctuations in a poloidal plane. The contours of regions that meet the blob conditions defined above are highlighted, with the center of mass of each blob identified. Figure~\ref{Fig:blob_detection} reveals that triangularity affects the blob properties. Blobs in PT plasmas are larger compared to those in NT plasmas. Furthermore, PT plasmas exhibit blobs across the entire LFS region, some of them extending towards the divertor region. In contrast, blobs in NT plasmas are mostly localized near the OMP. This observation confirms the reduced blob transport to the first wall in NT plasmas, as experimentally observed in TCV discharges \cite{Han2021}.

\begin{table}[H]
\centering
\caption{Blob detection results for both PT and NT plasmas with the average number of blobs, average radius, and average velocity. The radius and velocity are normalized to $\rho_{s0}$ and $c_{s0}$, respectively.}
\setlength{\heavyrulewidth}{0.4mm} 
\setlength{\lightrulewidth}{0.4mm} 
\begin{tabular}{l >{\centering\arraybackslash}p{1cm} >{\centering\arraybackslash}p{1cm} | >{\centering\arraybackslash}p{1cm} >{\centering\arraybackslash}p{1cm} | >{\centering\arraybackslash}p{1cm} >{\centering\arraybackslash}p{1cm} | >{\centering\arraybackslash}p{1cm} >{\centering\arraybackslash}p{1cm} | >{\centering\arraybackslash}p{1cm} >{\centering\arraybackslash}p{1cm} | >{\centering\arraybackslash}p{1cm} >{\centering\arraybackslash}p{1cm}}
\toprule
& \multicolumn{2}{c}{$\nu_0=0.1$} & \multicolumn{2}{c}{$\nu_0=0.3$} & \multicolumn{2}{c}{$\nu_0=1.0$} & \multicolumn{2}{c}{$s_{T0}=0.05$} & \multicolumn{2}{c}{$s_{T0}=0.15$} & \multicolumn{2}{c}{$s_{T0}=0.3$} \\
\cmidrule(lr){2-3} \cmidrule(lr){4-5} \cmidrule(lr){6-7} \cmidrule(lr){8-9} \cmidrule(lr){10-11} \cmidrule(lr){12-13}
& PT & NT & PT & NT & PT & NT & PT & NT & PT & NT & PT & NT \\
\midrule
\# Blobs & 37 & 39 & 31 & 34 & 30 & 34 & 25 & 35 & 31 & 34 & 27 & 45 \\
Radius $[\rho_{s0}]$& 7.66 & 7.03 & 8.89 & 7.12 & 9.96 & 7.85 & 9.95 & 7.69 & 8.89 & 7.12 & 7.01 & 6.21 \\
Velocity $[c_{s0}]$ & 0.07 & 0.05 & 0.12 & 0.10 & 0.15 & 0.10 & 0.12 & 0.09 & 0.12 & 0.10 & 0.16 & 0.07 \\
\bottomrule
\label{Blob_table}
\end{tabular}
\end{table}

The results of the blob detection methods applied over a period of $10 t_0$ are summarized in Table~\ref{Blob_table}, detailing the average number of blobs, their average radius, and their average radial velocity. Confirming the findings in Figure~\ref{Fig:blob_detection}, PT plasmas exhibit larger blob sizes and higher propagation velocities compared to NT plasmas. Blob size and propagation velocities are found to follow the intensity of background turbulence; more precisely these quantities increase with $\nu_0$ and decrease with $s_{T0}$. The effect of triangularity affects the blob velocity more significantly than their size, with NT plasmas exhibiting a greater number of blobs. A detailed analysis of these observations is provided below.

%% file: 6-2_Two_region.tex
\subsection{Two-region model for blob analysis}
A large number of theoretical investigations have focused on the understanding of blob dynamics (see, e.g., Refs.~\cite{Krasheninnikov2001, Yu2003, Aydemir2005, Ippolito2011}). An estimate of the radial velocity of blobs in different regimes was first presented in Ref.~\cite{Myra2006}, providing an analogy with an electrical circuit. The regimes are identified depending on the current restoring the quasi-neutrality, which balances the magnetic curvature and gradient drift responsible for the charge separation driving the $E \times B$ radial motion. This analysis identifies four different regimes of blob motion, defined by two parameters: the normalized cross-section size $\Theta$, and the effective collisionality parameter $\Lambda$. These parameters are defined as follows: 
\begin{align}
    \Theta =\hat{a}^{5/2},
\end{align}
and
\begin{align}
    \Lambda = \frac{\nu_{ei} L_1^2}{\rho_s \Omega_{ce} L_2}.
\end{align}
Here, $\nu_{ei}$ is the electron-to-ion collision frequency at the OMP, $L_1$ represents the connection length from the OMP to the X-point, $L_2$ is the connection length from the X-point to the outer target, and $\hat{a}$ is the normalized radial blob size defined as follows:
\begin{align}
    \hat{a}=\frac{a_b R^{1/5}}{L_\parallel^{2/5}\rho_s^{4/5}},
\label{normalized_a}
\end{align}
where $a_b$ is the blob size and $L_\parallel$ is the parallel connection length to the outer divertor target.

The four different regimes are: (i) the \textit{sheath-connected regime} $(C_s)$, which occurs when plasma collisionality is low ($\Lambda <1$) and blobs extend to the outer target, with the charge separation balanced by the current flowing through the sheath; (ii) the \textit{connected ideal-interchange regime} $(C_i)$, also found at low plasma collisionality, where the drive is balanced by the ion polarization current in the divertor region; (iii) the \textit{resistive X-point regime} (RX), characterized by high plasma collisionality ($\Lambda >1$) and parallel current flowing between the OMP and the divertor regions; and, finally, (iv) the \textit{resistive ballooning} (RB) regime, where the filaments are disconnected from the divertor regions due to high collisionality, with the ion polarization current balancing the drive. Each regime exhibits a distinct velocity scaling that are separated in terms of the normalized radial blob size, $\hat{a}$, defined in Eq.~(\ref{normalized_a}), and the normalized blob velocity,
\begin{align}
    \hat{v}= \frac{v_R}{ ({2 L_\parallel \rho_s^2/R^3})^{1/5} c_s},
\label{normalized_blobs}
\end{align}
where $v_R$ is the radial velocity of blobs \cite{Aydemir2005, Yu2006}.

\begin{figure}[H]
\begin{center}
\includegraphics[width=\textwidth]{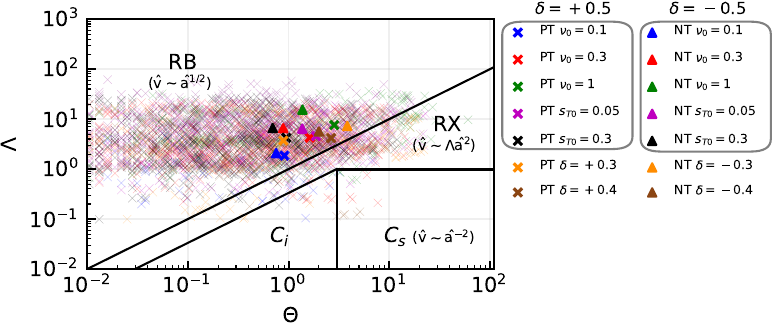}
\caption{Identification of four different regimes of blob dynamics in the ($\Theta,\Lambda$) diagram. Individual blobs detected in the simulations are shown with semi-transparent markers in the background, while the average values are indicated by solid markers.}
\label{Fig:Two_region_model1}
\end{center}
\end{figure}

Figure~\ref{Fig:Two_region_model1} illustrates the result of the regime identification for the detected blobs in the $(\Theta, \Lambda)$ diagram. Blobs are predominantly found in the RB regime, corresponding to L-mode plasma conditions with high plasma resistivity $\nu_0$, resulting in high values of $\Lambda$. These observations are consistent with those in TCV experiments, where most blobs are found in the RB regime \cite{Tsui2018}. Increasing plasma resistivity tends to increase both $\Lambda$ and $\Theta$, whereas increasing input heating power tends to decrease both, aligning with the experimental observations from the TCV tokamak \cite{Offeddu2022}.

The differences between NT and PT plasmas are also evident in Figure~\ref{Fig:Two_region_model1}. First, NT plasmas exhibit larger values of $\Lambda$, pointing out that blobs are less connected to the sheath compared to the PT case. This is further demonstrated in Figure~\ref{Fig:blob_detection}, where no blobs extend to the divertor region, in contrast to PT plasmas, where a few are detected in this region. Second, PT plasmas present larger values of $\Theta$, indicating larger sizes. The main physical mechanism driving the differences between NT and PT plasmas can be explained by the two-region model \cite{Myra2006}, which is the main focus of Section~\ref{Sec3} \textcolor{blue}{C}.

%% file: 6-3_Scaling_law.tex
\subsection{Derivation of the analytical blob scaling law including plasma shaping effects}
The previous results highlight the effect of triangularity on blob properties, indicating that NT plasmas are characterized by smaller and slower blobs but are more numerous compared to PT plasmas. The physical mechanisms behind these differences can be understood based on the reduction of the interchange drive in NT configurations, similar to the reduction of the $L_p$ described in Section~\ref{Sec3}. 

Following the analytical derivation in Ref.~\cite{Paruta2019}, we start from the GBS density and vorticity equations, Eqs.~(\ref{GBS_density}) and (\ref{GBS_vorticity}), to deduce the blob properties. It follows that:
\begin{align}
    \frac{\partial \omega_1}{\partial t} + \frac{R_0}{\rho_{s0}}[\phi_1, \omega_1] &= \frac{1}{n_1}\nabla_\parallel j_{\parallel 1} + \frac{2 T_{1}}{n_1} \mathcal{C}(n_1),\label{model_eq1} \\
    \frac{\partial n_1}{\partial t} + \frac{R_0}{\rho_{s0}}[\phi_1, n_1] &=0 ,\label{model_eq2}\\
    \frac{\partial \omega_2}{\partial t} + \frac{R_0}{\rho_{s0}}[\phi_2, \omega_2] &= \frac{1}{n_2}\nabla_\parallel j_{\parallel 2},\label{model_eq3}\\
    \frac{\partial n_2}{\partial t} + \frac{R_0}{\rho_{s0}}[\phi_2, n_2] &=0,\label{model_eq4} 
\end{align}
where the indices 1 and 2 represent the OMP and target regions, and $\mathcal{C}$ is the curvature operator at the outer midplane. In the density equation, the parallel gradient and magnetic curvature and gradient terms are neglected because they are smaller compared to the $\bm{E} \times \bm{B}$ drift term. In the vorticity equation, the parallel gradient terms associated with the polarization current are also neglected. Furthermore, the curvature operator is neglected in the target region. In Eqs.~(\ref{model_eq1}--\ref{model_eq4}), the curvature operator $\mathcal{C}$ in the upstream region represents the driving mechanism for charge separation. 

To make analytical progress, we replace the curvature operator in Eq.~(\ref{model_eq1}) with Eq.~(\ref{Curvature_operator}), and the Poisson bracket terms are rewritten as advective $\mathbf{E} \times \mathbf{B}$ terms, $[\phi, \omega] = \mathbf{v}_E \cdot \nabla \omega$. The two-region model outlined in Eqs.~(\ref{model_eq1}--\ref{model_eq4}) can then be rewritten as follows:

\begin{align}
    &\bigg(\frac{\partial}{\partial t} + \frac{R_0}{\rho_{s0}}\mathbf{v}_{E,1} \cdot \nabla \bigg) \nabla_\perp^2 \phi_1 = \frac{1}{\nu L_1^2} \frac{\phi_1 - \phi_2}{n_1} + \frac{2\rho_s^2}{n_{1}}\frac{\partial n_1}{\partial Z} \mathcal{C}(q, \kappa, \delta), \label{scaling1}\\
    &\bigg( \frac{\partial}{\partial t} + \rho_*^{-1} \mathbf{v}_{E,1} \cdot \nabla n_1 \bigg) =0, \\
    &\bigg(\frac{\partial}{\partial t} + \rho_*^{-1} \textbf{v}_{E,2} \cdot \nabla \bigg) \nabla_\perp^2 \phi =-\frac{1}{\nu L_1 L_2} \frac{\phi_1 - \phi_2}{n_2} + \frac{\phi_2 - \phi_f}{\rho_s L_2}, \\
    &\bigg(\frac{\partial}{\partial t} + \rho_*^{-1} \mathbf{v}_{E,2} \cdot \nabla  \bigg) n_2 =0. \label{scaling4}
\end{align}
Eqs.~(\ref{scaling1}--\ref{scaling4}) are identical to Eqs.~(A5--A8) in Appendix A of Ref.~\cite{Paruta2019}, except that the curvature operator is now replaced by Eq.~(\ref{Curvature_operator}) to include plasma shaping parameters in the blob analysis, which allow us to derive scaling laws for blob size and velocity that account for the effects of $\delta$ on the blob dynamics. 

Since the derivation procedure of the scaling of blob size and velocity is identical to the one reported in Appendix A of Ref.~\cite{Paruta2019}, we directly present the final analytical scaling laws for these parameters: 
\begin{align}
    a^* & \propto \mathcal{C}(q, \delta, \kappa)^{1/5}a^*_{\text{ref}}, \label{size_scaling}\\
    v^* & \propto \mathcal{C}(q, \delta, \kappa)^{1/2}v^*_{\text{ref}}, \label{velocity_scaling}
\end{align}
where $a^{*}_{\textrm{ref}}$ and $v^{*}_{\textrm{ref}}$ are the reference scaling law for blob size and velocity, defined in Eqs.~(A26) and (A29) of Ref.~\cite{Paruta2019}, as follows: 
\begin{align}
    a^*_{\text{ref}} &= \bigg({2 \rho_s^4 L_2^2} \frac{\Delta n_1}{\overbar{n}_{1}} \rho_*^{-1}\bigg)^{1/5}, \label{paruta_size_scaling} \\
    v^*_{\text{ref}}&= \rho_s \bigg(\frac{\pi}{4} \frac{\delta n_1^5}{\Delta n_1^2 \overbar{n}^3_{1}}\rho_s^2 L_2 \rho_*^2 \bigg)^{1/5}. \label{paruta_velocity_scaling}
\end{align}
These expressions are derived under the assumptions $\kappa=1$ and $\delta=0$, where $\Delta n$ represents an estimate of the radial variation of the blob density. Eqs.~(\ref{size_scaling}--\ref{velocity_scaling}) extend the scaling laws in Eqs.~(\ref{paruta_size_scaling}--\ref{paruta_velocity_scaling}) to include the effects of plasma shaping, revealing that plasma shaping is expected to have a more significant impact on velocity than on size. This conclusion is consistent with the findings in Table~\ref{Blob_table}, where variations in velocity are more pronounced than differences in blob size when triangularity is varied. 

\begin{figure}[H]
\begin{center}
\includegraphics[width=\textwidth]{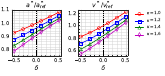}
\caption{Theoretical scaling laws for blob size and radial velocity, derived in Eqs.~(\ref{size_scaling}--\ref{velocity_scaling}) as a functions of $\delta$, for different values of $\kappa$. The values of size and radial velocity are normalized to the reference scaling laws for blob size, $a^*_{\text{ref}}$, and blob velocity, $v^*_{\text{ref}}$, defined in Eqs.~(\ref{paruta_size_scaling}--\ref{paruta_velocity_scaling}), for the case with $\kappa=1$ and $\delta=0$.}
\label{Fig:Blob_scaling}
\end{center}
\end{figure}

The analytical scaling law derived in Eqs.~(\ref{size_scaling}--\ref{velocity_scaling}) qualitatively reproduces the trend observed in nonlinear GBS simulations, as shown in Figure~\ref{Fig:Blob_scaling}.  The effect of triangularity becomes more pronounced as $\kappa$ increases. In the present simulations with $\kappa=1.6$, the analytical scaling law predicts more than a 40\% difference in blob velocity and more than a 20\% difference in size between NT ($\delta=-0.5$) and PT ($\delta=+0.5$) discharges. Along with the nonlinear blob analysis in Table~\ref{Blob_table}, these observations support that NT plasmas benefit from mitigated first-wall erosion due to their reduced blob size and slower propagation velocity.

%% file: 7_Conclusion.tex
\section{Conclusion}\label{Sec7}
The effects of NT on SOL plasma turbulence and blob dynamics in L-mode DN plasmas are investigated using global, nonlinear, three-dimensional, flux-driven, two-fluid GBS simulations. To explore the impact of triangularity under a range of conditions, we perform a parameter scan, varying triangularity, plasma resistivity, and input heating power across both NT and PT plasmas.

First, consistent with previous studies in SN configurations \cite{Lim2023}, NT configurations exhibit reduced SOL plasma turbulence with respect to PT discharges. The physical mechanism underlying this reduction is the same as the one observed in SN configurations; i.e., the reduction of interchange drive in the bad curvature region. The comparison of the analytical $L_p$ estimate with nonlinear GBS simulations confirms a smaller $L_p$ in the near SOL region.

Second, our investigations of the heat load at the targets and the power-sharing asymmetry consistently show that NT plasmas exhibit mitigated heat load at the outer targets across all simulations, while a non-negligible amount of heat flux is directed to the inner targets, thereby reducing the in-out power asymmetry. The enhanced inner target heat load in NT plasmas results from both the longer separatrix at the HFS and the reduced cross-field transport at the LFS. The comparison of the up-down power-sharing asymmetry between the predictive power-scaling law derived in Ref.~\cite{Lim2024} and the numerical simulations confirms that NT plasmas display reduced up-down asymmetry at the outer targets compared to PT plasmas. This reduction is due to the non-negligible inner heat loads and the decreased heat flux crossing the separatrix in the LFS region.

Finally, a blob detection technique is used to investigate the effect of triangularity on the blob dynamics. The analysis reveals that NT plasmas feature smaller sizes and slower propagation velocities, which are associated with the reduced RBMs in the SOL region. Applying the two-region model to the blobs detected from nonlinear simulations shows that the majority of blobs belong to the RB regime. By introducing elongation and triangularity effects in the two-region model for blob motion, we generalize the analytical scaling law previously derived in Ref.~\cite{Paruta2019}, obtaining qualitative agreements with the findings from nonlinear GBS simulations.

Overall, NT plasmas in DN configurations exhibit several advantages with respect to PT plasmas, including suppressed plasma fluctuation, mitigated power loads at the outer targets, reduced power-sharing asymmetry, and decreased wall interactions with blobs. While these findings highlight the benefits of using NT plasmas in L-mode operation compared to conventional PT plasmas in H-mode scenarios, they underscore the challenges posed by the reduced $L_p$ in NT scenarios. To further evaluate the viability of NT plasmas in DN configurations for power handling, our next steps involve achieving plasma detachment in these configurations and exploring more realistic wall geometries, such as baffled divertors.

%% file: 0_Main.bbl
\begin{thebibliography}{10}

\bibitem{Soukhanovskii2013}
V.A. Soukhanovskii, R.E. Bell, A.~Diallo, S.~Gerhardt, S.~Kaye, E.~Kolemen, B.P. LeBlanc, A.~McLean, J.E. Menard, S.F. Paul, M.~Podesta, R.~Raman, D.D. Ryutov, F.~Scotti, R.~Kaita, R.~Maingi, D.M. Mueller, A.L. Roquemore, H.~Reimerdes, G.P. Canal, B.~Labit, W.~Vijvers, S.~Coda, B.P. Duval, T.~Morgan, J.~Zielinski, G.~{De Temmerman}, and B.~Tal.
\newblock Advanced divertor configurations with large flux expansion.
\newblock {\em J. Nucl. Mater.}, \href{https://doi.org/10.1016/j.jnucmat.2013.01.015}{\textbf{438} S96-S101}, (2013).

\bibitem{Reimerdes2020}
H.~Reimerdes et~al.
\newblock Assessment of alternative divertor configurations as an exhaust solution for {DEMO}.
\newblock {\em Nucl. Fusion}, \href{https://doi.org/10.1088/1741-4326/ab8a6a}{\textbf{60} 066030}, (2020).

\bibitem{Militello2021}
F.~Militello et~al.
\newblock {Preliminary analysis of alternative divertors for DEMO}.
\newblock {\em Nucl. Mater. Energy}, \href{https://doi.org/10.1016/j.nme.2021.100908}{\textbf{26} 100908}, (2021).

\bibitem{Wenninger2018}
R.~Wenninger et~al.
\newblock {Power Handling and Plasma Protection Aspects that affect the design of the DEMO divertor and first wall}.
\newblock {\em (IAEA-CN--234) IAEA Conference Proceedings}, (2018).

\bibitem{Brunner2018}
D.~Brunner, A.Q. Kuang, B.~LaBombard, and J.L. Terry.
\newblock {The dependence of divertor power sharing on magnetic flux balance in near double-null configurations on Alcator C-Mod}.
\newblock {\em Nucl. Fusion}, \href{https://doi.org/10.1088/1741-4326/aac006}{\textbf{58} 076010}, (2018).

\bibitem{Meyer2005}
H.~Meyer et~al.
\newblock {H-mode physics of near double null plasmas in {MAST} and its applications to other tokamaks}.
\newblock {\em Nucl. Fusion}, \href{https://doi.org/10.1088/0029-5515/46/1/008}{\textbf{46} 64}, (2005).

\bibitem{Smick2013}
N.~Smick, B.~LaBombard, and I.H. Hutchinson.
\newblock {Transport and drift-driven plasma flow components in the Alcator C-Mod boundary plasma}.
\newblock {\em Nucl. Fusion}, \href{https://doi.org/10.1088/0029-5515/53/2/023001}{\textbf{53} 023001}, (2013).

\bibitem{LaBombard2017}
B.~LaBombard et~al.
\newblock High-field side scrape-off layer investigation: Plasma profiles and impurity screening behavior in near-double-null configurations.
\newblock {\em Nucl. Mater. Energy}, \href{https://doi.org/10.1016/j.nme.2016.10.006}{{12} 139}, (2017).

\bibitem{Kikuchi2015}
M.~Kikuchi et~al.
\newblock Perspective of negative triangularity tokamak as fusion energy system.
\newblock {\em 42nd European Physical Society Conference on Plasma Physics, EPS 2015 (Lisbon, Portugal, 22-26 June 2015)}, \href{http://ocs.ciemat.es/EPS2015PAP/pdf/P4.179.pdf}{\textbf{3} P4.179}, (2015).

\bibitem{Austin2019}
{Austin, M. E. and Marinoni, A. and Walker, M. L. and Brookman, M. W. and deGrassie, J. S. and Hyatt, A. W. and McKee, G. R. and Petty, C. C. and Rhodes, T. L. and Smith, S. P. and Sung, C. and Thome, K. E. and Turnbull, A. D.}
\newblock {Achievement of Reactor-Relevant Performance in Negative Triangularity Shape in the DIII-D Tokamak}.
\newblock {\em Phys. Rev. Lett.}, \href{https://doi.org/10.1103/PhysRevLett.122.115001}{\textbf{122} 115001}, (2019).

\bibitem{Coda2022}
S~Coda, A~Merle, O~Sauter, L~Porte, F~Bagnato, J~Boedo, T~Bolzonella, O~Février, B~Labit, A~Marinoni, A~Pau, L~Pigatto, U~Sheikh, C~Tsui, M~Vallar, T~Vu, and The~TCV Team.
\newblock {Enhanced confinement in diverted negative-triangularity L-mode plasmas in TCV}.
\newblock {\em Plasma Phys. Control. Fusion}, \href{https://doi.org/10.1088/1361-6587/ac3fec}{\textbf{64} 014004}, (2021).

\bibitem{Marinoni2019}
A.~Marinoni, M.~E. Austin, A.~W. Hyatt, M.~L. Walker, J.~Candy, C.~Chrystal, C.~J. Lasnier, G.~R. McKee, T.~Odstrčil, C.~C. Petty, M.~Porkolab, J.~C. Rost, O.~Sauter, S.~P. Smith, G.~M. Staebler, C.~Sung, K.~E. Thome, A.~D. Turnbull, L.~Zeng, and DIII-D Team.
\newblock {H-mode grade confinement in L-mode edge plasmas at negative triangularity on DIII-D}.
\newblock {\em Phys. Plasmas}, \href{https://doi.org/10.1063/1.5091802}{\textbf{26} 042515}, (2019).

\bibitem{Happel2023}
{T. Happel and T. Pütterich and D. Told and M. Dunne and R. Fischer and J. Hobirk and R.M. McDermott and U. Plank and ASDEX Upgrade Team the}.
\newblock {Overview of initial negative triangularity plasma studies on the ASDEX Upgrade tokamak}.
\newblock {\em Nucl. Fusion}, \href{https://dx.doi.org/10.1088/1741-4326/ac8563}{\textbf{63} 016002}, (2023).

\bibitem{Balestri2024}
A~Balestri, P~Mantica, A~Mariani, F~Bagnato, T~Bolzonella, J~Ball, S~Coda, M~Dunne, M~Faitsch, P~Innocente, P~Muscente, O~Sauter, M~Vallar, E~Viezzer, the TCV~Team, and the EUROfusion Tokamak Exploitation~Team.
\newblock {Experiments and gyrokinetic simulations of TCV plasmas with negative triangularity in view of DTT operations}.
\newblock {\em Plasma Phys. Control. Fusion}, \href{https://dx.doi.org/10.1088/1361-6587/ad4674}{\textbf{66} 065031}, (2024).

\bibitem{Mariani2024}
A.~Mariani, A.~Balestri, P.~Mantica, G.~Merlo, R.~Ambrosino, L.~Balbinot, D.~Brioschi, I.~Casiraghi, A.~Castaldo, L.~Frassinetti, V.~Fusco, P.~Innocente, O.~Sauter, and G.~Vlad.
\newblock First-principle based predictions of the effects of negative triangularity on dtt scenarios.
\newblock {\em Nucl. Fusion}, \href{https://dx.doi.org/10.1088/1741-4326/ad2abc}{\textbf{64} 046018}, (2024).

\bibitem{Marinoni2024}
A.~Marinoni, M.~E. Austin, J.~Candy, C.~Chrystal, S.~R. Haskey, M.~Porkolab, J.~C. Rost, and F.~Scotti.
\newblock {Nonlinear gyrokinetic modelling of high confinement negative triangularity plasmas}.
\newblock {\em Nucl. Fusion}, \href{https://dx.doi.org/10.1088/1741-4326/ad5a1c}{\textbf{64} 086045}, (2024).

\bibitem{DiGiannatale2024}
Giovanni~Di Giannatale, Alberto Bottino, Stephan Brunner, Moahan Murugappan, and Laurent Villard.
\newblock {System size scaling of triangularity effects on global temperature gradient-driven gyrokinetic simulations}.
\newblock {\em Plasma Phys. Control. Fusion}, \href{https://dx.doi.org/10.1088/1361-6587/ad5df9}{\textbf{66} 095003}, (2024).

\bibitem{Riva2017}
F.~Riva et~al.
\newblock Plasma shaping effects on tokamak scrape-off layer turbulence.
\newblock {\em Plasma Phys. Control. Fusion}, \href{https://doi.org/10.1088/1361-6587/aa5322}{\textbf{59} 035001}, (2017).

\bibitem{Muscente2023}
P.~Muscente, P.~Innocente, J.~Ball, and S.~Gorno.
\newblock Analysis of edge transport in l-mode negative triangularity tcv discharges.
\newblock {\em Nucl. Mater. Energy}, \href{https://doi.org/10.1016/j.nme.2023.101386}{\textbf{34} 101386}, (2023).

\bibitem{Lim2023}
K.~Lim, M.~Giacomin, P.~Ricci, A.~Coelho, O.~Février, D.~Mancini, D.~Silvagni, and L.~Stenger.
\newblock {Effect of triangularity on plasma turbulence and the SOL-width scaling in L-mode diverted tokamak configurations}.
\newblock {\em Plasma Phys. Control. Fusion}, \href{https://dx.doi.org/10.1088/1361-6587/acdc52}{\textbf{65} 085006}, (2023).

\bibitem{Tonello2024}
E~Tonello, F~Mombelli, O~Février, G~Alberti, T~Bolzonella, G~Durr-Legoupil-Nicoud, S~Gorno, H~Reimerdes, C~Theiler, N~Vianello, M~Passoni, the TCV~Team, and the WPTE~Team.
\newblock {Modelling of power exhaust in TCV positive and negative triangularity L-mode plasmas}.
\newblock {\em Plasma Phys. Control. Fusion}, \href{https://dx.doi.org/10.1088/1361-6587/ad3c19}{\textbf{66} 065006}, (2024).

\bibitem{Lim2024}
K.~Lim, P.~Ricci, L.~Stenger, B.~De Lucca, G.~Durr-Legoupil-Nicoud, O.~Février, C.~Theiler, and K.~Verhaegh.
\newblock {Predictive power-sharing scaling law in double-null L-mode plasmas}.
\newblock {\em Nucl. Fusion}, \href{https://dx.doi.org/10.1088/1741-4326/ad7743}{\textbf{64} 106057}, (2024).

\bibitem{Zeiler1997}
A.~Zeiler et~al.
\newblock {Nonlinear reduced Braginskii equations with ion thermal dynamics in toroidal plasma}.
\newblock {\em Phys. Plasams}, \href{https://doi.org/10.1063/1.872368} {\textbf{4} 2134}, (1997).

\bibitem{Ricci2012}
P.~Ricci, F.~D. Halpern, S.~Jolliet, J.~Loizu, A.~Mosetto, A.~Fasoli, I.~Furno, and C.~Theiler.
\newblock Simulation of plasma turbulence in scrape-off layer conditions: the gbs code, simulation results and code validation.
\newblock {\em Plasma Phys. Control. Fusion}, \href{https://dx.doi.org/10.1088/0741-3335/54/12/124047}{\textbf{54} 124047}, (2012).

\bibitem{Giacomin2021}
M.~Giacomin et~al.
\newblock {The GBS code for the self-consistent simulation of plasma turbulence and kinetic neutral dynamics in the tokamak boundary}.
\newblock {\em J. Comput. Phys.}, \href{https://doi.org/10.1016/j.jcp.2022.111294}{\textbf{463} 111294}, (2022).

\bibitem{Giacomin2020_2}
M.~Giacomin, L.N. Stenger, and P.~Ricci.
\newblock Turbulence and flows in the plasma boundary of snowflake magnetic configurations.
\newblock {\em Nucl. Fusion}, \href{https://dx.doi.org/10.1088/1741-4326/ab6435}{\textbf{60} 024001}, (2020).

\bibitem{Beadle2020}
C.~Beadle and P~Ricci.
\newblock Understanding the turbulent mechanisms setting the density decay length in the tokamak scrape-off layer.
\newblock {\em J. Plasma Phys.}, \href{https://doi.org/10.1017/S0022377820000094}{\textbf{86} 175860101}, (2020).

\bibitem{Coelho2022}
A.~Coelho, J.~Loizu, P.~Ricci, and M.~Giacomin.
\newblock Global fluid simulation of plasma turbulence in a stellarator with an island divertor.
\newblock {\em Nucl. Fusion}, (2022).

\bibitem{Loizu2012}
J.~Loizu, P.~Ricci, F.~D. Halpern, and S.~Jolliet.
\newblock {Boundary conditions for plasma fluid models at the magnetic presheath entrance}.
\newblock {\em Phys. Plasmas}, \href{https://doi.org/10.1063/1.4771573}{\textbf{19} 122307}, (2012).

\bibitem{Reimerdes2022}
H.~Reimerdes et~al.
\newblock {Overview of the TCV tokamak experimental programme}.
\newblock {\em Nucl. Fusion}, \href{https://dx.doi.org/10.1088/1741-4326/ac369b}{\textbf{62} 042018}, (2022).

\bibitem{Fundamenski2007}
W.~Fundamenski, O.E. Garcia, V.~Naulin, R.A. Pitts, A.H. Nielsen, J.~Juul Rasmussen, J.~Horacek, J.P. Graves, and JET~EFDA contributors.
\newblock {Dissipative processes in interchange driven scrape-off layer turbulence}.
\newblock {\em Nucl. Fusion}, \href{https://dx.doi.org/10.1088/0029-5515/47/5/006}{\textbf{47} 417}, 2007.

\bibitem{Mosetto2013}
A.~Mosetto et~al.
\newblock Turbulent regimes in the tokamak scrape-off layer.
\newblock {\em Phys. Plasmas}, \href{https://doi.org/10.1063/1.4821597}{\textbf{20} 092308}, (2013).

\bibitem{Giacomin2020}
M.~Giacomin and P.~Ricci.
\newblock {Investigation of turbulent transport regimes in the tokamak edge by using two-fluid simulations}.
\newblock {\em J. Plasma Phys.}, \href{https://doi:10.1017/S0022377820000914}{\textbf{5} 86}, (2020).

\bibitem{Kikuchi2019}
M.~Kikuchi, T.~Takizuka, S.~Medvedev, T.~Ando, D.~Chen, J.X. Li, M.~Austin, O.~Sauter, L.~Villard, A.~Merle, M.~Fontana, Y.~Kishimoto, and K.~Imadera.
\newblock L-mode-edge negative triangularity tokamak reactor.
\newblock {\em Nucl. Fusion}, \href{https://dx.doi.org/10.1088/1741-4326/ab076d}{\textbf{59} 056017}, (2019).

\bibitem{Giacomin2021_2}
M.~Giacomin et~al.
\newblock {Theory-based scaling laws of near and far scrape-off layer widths in single-null L-mode discharges}.
\newblock {\em Nucl. Fusion}, \href{https://doi.org/10.1088/1741-4326/abf8f6}{\textbf{61} 76002}, (2021).

\bibitem{Ricci2008}
P.~Ricci, B.~N. Rogers, and S.~Brunner.
\newblock High- and low-confinement modes in simple magnetized toroidal plasmas.
\newblock {\em Phys. Rev. Lett.}, \href{https://doi.org/10.1103/PhysRevLett.100.225002}{\textbf{100} 225002}, (2008).

\bibitem{Ricci2013}
P.~Ricci and B.~N. Rogers.
\newblock Plasma turbulence in the scrape-off layer of tokamak devices.
\newblock {\em Phys. Plasmas}, \href{https://doi.org/10.1063/1.4789551}{\textbf{20} 010702}, (2013).

\bibitem{Liu2014}
S.~C. Liu, H.~Y. Guo, L.~Wang, H.~Q. Wang, K.~F. Gan, T.~Y. Xia, G.~S. Xu, X.~Q. Xu, Z.~X. Liu, L.~Chen, N.~Yan, W.~Zhang, R.~Chen, L.~M. Shao, S.~Ding, G.~H. Hu, Y.~L. Liu, N.~Zhao, Y.~L. Li, X.~Z. Gong, and X.~Gao.
\newblock {Effects of heating power on divertor in-out asymmetry and scrape-off layer flow in reversed field on Experimental Advanced Superconducting Tokamak}.
\newblock {\em Phys. Plasmas}, \href{https://doi.org/10.1063/1.4904205}{\textbf{21} 122514}, (2014).

\bibitem{Liu2016}
J.B. Liu, H.Y. Guo, L.~Wang, G.S. Xu, T.Y. Xia, S.C. Liu, X.Q. Xu, Jie Li, L.~Chen, N.~Yan, H.Q. Wang, J.C. Xu, W.~Feng, L.M. Shao, G.Z. Deng, H.~Liu, and EAST~Probe Team.
\newblock {In–out asymmetry of divertor particle flux in H-mode with edge localized modes on EAST}.
\newblock {\em Nucl Fusion}, \href{https://dx.doi.org/10.1088/0029-5515/56/6/066006}{\textbf{56} 066006}, (2016).

\bibitem{Du2017}
H.~Du et~al.
\newblock {Role of $E \times B$ on in-out divertor asymmetry in high recycling/partial detachment regimes under L-mode and H-mode conditions}.
\newblock {\em Nucl. Fusion}, \href{https://10.1088/1741-4326/aa7d79}{\textbf{57} 116022}, (2017).

\bibitem{Morel1999}
K.M Morel, G.F Counsell, and P~Helander.
\newblock {Asymmetries in the divertor power loading in START}.
\newblock {\em J. Nucl. Mater.}, \href{https://doi.org/10.1016/S0022-3115(98)00851-4}{\textbf{266} 1040}, (1999).

\bibitem{Petrie2001}
T.W. Petrie et~al.
\newblock {The effect of divertor magnetic balance on H-mode performance in DIII-D}.
\newblock {\em J. Nucl. Mater.}, \href{https://doi.org/10.1016/S0022-3115(00)00492-X}{\textbf{290} 935}, (2001).

\bibitem{Temmerman2011}
G.~{De Temmerman}, A.~Kirk, E.~Nardon, and P.~Tamain.
\newblock Heat load asymmetries in {MAST}.
\newblock {\em J. Nucl. Mater.}, \href{https://doi.org/10.1016/j.jnucmat.2010.10.003}{\textbf{415} S383}, (2011).

\bibitem{Fevrier2021}
O.~F\'evrier et~al.
\newblock {Detachment in conventional and advanced double-null plasmas in TCV}.
\newblock {\em Nucl. Fusion}, \href{https://doi.org/10.1088/1741-4326/ac27c6}{\textbf{61} 116064}, (2021).

\bibitem{Rognlien1999}
T.D. Rognlien, G.D. Porter, and D.D. Ryutov.
\newblock {Influence of $E \times B$ and $\nabla B$ drift terms in 2D edge/SOL transport simulations}.
\newblock {\em J. Nucl. Mater.}, \href{https://doi.org/10.1016/S0022-3115(98)00835-6}{\textbf{266} 654}, (1999).

\bibitem{Cohen1999}
R.~H. Cohen and D.~Ryutov.
\newblock Drifts, boundary conditions and convection on open field lines.
\newblock {\em Phys. Plasmas}, \href{https://doi.org/10.1063/1.873455}{\textbf{6} 1995}, (1999).

\bibitem{Rensink2000}
M.E. Rensink, S.L. Allen, G.D. Porter, and T.D. Rognlien.
\newblock {Simulation of Double-Null Divertor Plasmas with the UEDGE Code}.
\newblock {\em Contrib. Plasma. Phys}, \href{https://doi.org/10.1002/1521-3986(200006)40:3/4<302::AID-CTPP302>3.0.CO;2-L}{\textbf{40} 302-308}, (2000).

\bibitem{Rubino2020}
G.~Rubino et~al.
\newblock {Effect of drifts on SOL plasma in DTT Double Null configuration}.
\newblock {\em 47th EPS Conference on Plasma Physics}, \href{http://ocs.ciemat.es/EPS2021PAP/pdf/P3.1026.pdf}{\textbf{P3} 1026}, (2020).

\bibitem{Du2015}
H.~Du et~al.
\newblock {Effects of drifts and ballooning instability on the divertor in–out asymmetry in EAST tokamak}.
\newblock {\em J. Nucl. Materials}, \href{https://doi.org/10.1016/j.jnucmat.2014.12.093}{\textbf{463} 485}, (2015).

\bibitem{Schaffer1997}
M.J. Schaffer et~al.
\newblock {Pfirsch-Schluter currents in the JET divertor}.
\newblock {\em Nucl. Fusion}, \href{https://dx.doi.org/10.1088/0029-5515/37/1/I07}{\textbf{37} 83}, (1997).

\bibitem{Asakura2004}
N.~Asakura et~al.
\newblock {Driving mechanism of SOL plasma flow and effects on the divertor performance in {JT}-60U}.
\newblock {\em Nucl. Fusion}, \href{https://doi.org/10.1088/0029-5515/44/4/004}{\textbf{44} 503}, (2004).

\bibitem{Osawa2023}
R.T. Osawa, D.~Moulton, S.L. Newton, S.S. Henderson, B.~Lipschultz, and A.~Hudoba.
\newblock {SOLPS-ITER analysis of a proposed STEP double null geometry: impact of the degree of disconnection on power-sharing}.
\newblock {\em Nucl. Fusion}, \href{https://dx.doi.org/10.1088/1741-4326/acd863}{\textbf{63} 076032}, (2023).

\bibitem{Liu2012}
S.~C. Liu et~al.
\newblock {Divertor asymmetry and scrape-off layer flow in various divertor configurations in Experimental Advanced Superconducting Tokamak}.
\newblock {\em Phys. Plasmas}, \href{https://doi.org/10.1063/1.4707396}{\textbf{19} 042505}, (2012).

\bibitem{Myra2006_2}
J.~R. Myra, D.~A. D’Ippolito, D.~P. Stotler, S.~J. Zweben, B.~P. LeBlanc, J.~E. Menard, R.~J. Maqueda, and J.~Boedo.
\newblock {Blob birth and transport in the tokamak edge plasma: Analysis of imaging data}.
\newblock {\em Phys. Plasmas}, \href{https://doi.org/10.1063/1.2355668}{\textbf{13} 092509}, (2006).

\bibitem{Zweben2007}
S~J Zweben, J~A Boedo, O~Grulke, C~Hidalgo, B~LaBombard, R~J Maqueda, P~Scarin, and J~L Terry.
\newblock Edge turbulence measurements in toroidal fusion devices.
\newblock {\em Plasma Phys. Control. Fusion}, \href{https://dx.doi.org/10.1088/0741-3335/49/7/S01}{\textbf{49} S1}, (2007).

\bibitem{Ippolito2011}
D.~A. D’Ippolito, J.~R. Myra, and S.~J. Zweben.
\newblock {Convective transport by intermittent blob-filaments: Comparison of theory and experiment}.
\newblock {\em Phys. Plasmas}, \href{https://doi.org/10.1063/1.3594609}{\textbf{18} 060501}, (2011).

\bibitem{Rudakov2002}
D~L Rudakov, J~A Boedo, R~A Moyer, S~Krasheninnikov, A~W Leonard, M~A Mahdavi, G~R McKee, G~D Porter, P~C Stangeby, J~G Watkins, W~P West, D~G Whyte, and G~Antar.
\newblock {Fluctuation-driven transport in the DIII-D boundary}.
\newblock {\em Plasma Phys. Control. Fusion}, \href{https://dx.doi.org/10.1088/0741-3335/44/6/308}{\textbf{44} 717}, (2002).

\bibitem{Angus2012}
Justin~R. Angus, Sergei~I. Krasheninnikov, and Maxim~V. Umansky.
\newblock {Effects of parallel electron dynamics on plasma blob transport}.
\newblock {\em Phys. Plasmas}, \href{https://doi.org/10.1063/1.4747619}{\textbf{19} 082312}, (2012).

\bibitem{Walkden2013}
N~R Walkden, B~D Dudson, and G~Fishpool.
\newblock {Characterization of 3D filament dynamics in a MAST SOL flux tube geometry}.
\newblock {\em Plasma Phys. Control. Fusion}, \href{https://dx.doi.org/10.1088/0741-3335/55/10/105005}{\textbf{55} 105005}, (2013).

\bibitem{Easy2014}
L.~Easy, F.~Militello, J.~Omotani, B.~Dudson, E.~Havlíčková, P.~Tamain, V.~Naulin, and A.~H. Nielsen.
\newblock {Three dimensional simulations of plasma filaments in the scrape off layer: A comparison with models of reduced dimensionality}.
\newblock {\em Phys. Plasmas}, \href{https://doi.org/10.1063/1.4904207}{\textbf{21} 122515}, (2014).

\bibitem{Russell2004}
D.~A. Russell, D.~A. D'Ippolito, J.~R. Myra, W.~M. Nevins, and X.~Q. Xu.
\newblock {Blob Dynamics in 3D BOUT Simulations of Tokamak Edge Turbulence}.
\newblock {\em Phys. Rev. Lett.}, \href{https://dx.doi.org/10.1103/PhysRevLett.93.265001}{\textbf{93} 265001}, (2004).

\bibitem{Nespoli2017}
F~Nespoli, I~Furno, B~Labit, P~Ricci, F~Avino, F~D Halpern, F~Musil, and F~Riva.
\newblock {Blob properties in full-turbulence simulations of the TCV scrape-off layer}.
\newblock {\em Plasma Phys. Control. Fusion}, \href{https://dx.doi.org/10.1088/1361-6587/aa6276}{\textbf{59} 055009}, (2017).

\bibitem{Paruta2019}
Paola Paruta, C.~Beadle, P.~Ricci, and C.~Theiler.
\newblock {Blob velocity scaling in diverted tokamaks: A comparison between theory and simulation}.
\newblock {\em Phys. Plasmas}, \href{https://doi.org/10.1063/1.5080675}{\textbf{26} 032302}, (2019).

\bibitem{Ross2019}
A.~Ross, A.~Stegmeir, P.~Manz, D.~Groselj, W.~Zholobenko, D.~Coster, and F.~Jenko.
\newblock {On the nature of blob propagation and generation in the large plasma device: Global GRILLIX studies}.
\newblock {\em Phys. Plasmas}, \href{https://doi.org/10.1063/1.5095712}{\textbf{26} 102308}, (2019).

\bibitem{Zholobenko2023}
W.~Zholobenko, J.~Pfennig, A.~Stegmeir, T.~Body, P.~Ulbl, and F.~Jenko.
\newblock {Filamentary transport in global edge-SOL simulations of ASDEX Upgrade}.
\newblock {\em Nucl. Mater. Energy}, \href{https://doi.org/10.1016/j.nme.2022.101351}{\textbf{34} 101351}, (2023).

\bibitem{Myra2006}
J.~R. Myra, D.~A. Russell, and D.~A. D’Ippolito.
\newblock {Collisionality and magnetic geometry effects on tokamak edge turbulent transport. I. A two-region model with application to blobs}.
\newblock {\em Phys. Plasmas}, \href{https://doi.org/10.1063/1.2364858}{\textbf{13} 112502}, (2006).

\bibitem{scikit-image}
{S}t\'efan van~der Walt, {J}ohannes~{L}. {S}ch\"onberger, {J}uan {Nunez-Iglesias}, {F}ran\c{c}ois {B}oulogne, {J}oshua~{D}. {W}arner, {N}eil {Y}ager, {E}mmanuelle {G}ouillart, {T}ony {Y}u, and the scikit-image contributors.
\newblock scikit-image: image processing in {P}ython.
\newblock {\em PeerJ}, \href{https://doi.org/10.7717/peerj.453}{\textbf{2} e453}, (2014).

\bibitem{Han2021}
W.~Han et~al.
\newblock Suppression of first-wall interaction in negative triangularity plasmas on {TCV}.
\newblock {\em Nucl. Fusion}, \href{https://doi.org/10.1088/1741-4326/abdb95}{\textbf{61} 034001}, (2021).

\bibitem{Krasheninnikov2001}
S.I. Krasheninnikov.
\newblock On scrape off layer plasma transport.
\newblock {\em Physics Letters A}, \href{https://doi.org/10.1016/S0375-9601(01)00252-3}{\textbf{283} 368-370}, (2001).

\bibitem{Yu2003}
G.~Q. Yu and S.~I. Krasheninnikov.
\newblock {Dynamics of blobs in scrape-off-layer/shadow regions of tokamaks and linear devices}.
\newblock {\em Phys. Plasmas}, \href{https://doi.org/10.1063/1.1616937}{\textbf{10} 4413-4418}, (2003).

\bibitem{Aydemir2005}
A.~Y. Aydemir.
\newblock {Convective transport in the scrape-off layer of tokamaks}.
\newblock {\em Phys. Plasmas}, \href{https://doi.org/10.1063/1.1927539}{\textbf{12} 062503}, (2005).

\bibitem{Yu2006}
G.~Q. Yu, S.~I. Krasheninnikov, and P.~N. Guzdar.
\newblock {Two-dimensional modelling of blob dynamics in tokamak edge plasmas}.
\newblock {\em Phys. Plasmas}, \href{https://doi.org/10.1063/1.2193087}{\textbf{13} 042508}, (2006).

\bibitem{Tsui2018}
C.~K. Tsui, J.~A. Boedo, J.~R. Myra, B.~Duval, B.~Labit, C.~Theiler, N.~Vianello, W.~A.~J. Vijvers, H.~Reimerdes, S.~Coda, O.~Février, J.~R. Harrison, J.~Horacek, B.~Lipschultz, R.~Maurizio, F.~Nespoli, U.~Sheikh, K.~Verhaegh, N.~Walkden, TCV Team, and EUROfusion~MST1 Team.
\newblock {Filamentary velocity scaling validation in the TCV tokamak}.
\newblock {\em Phys. Plasmas}, \href{https://doi.org/10.1063/1.5038019}{\textbf{25} 072506}, (2018).

\bibitem{Offeddu2022}
N.~Offeddu, W.~Han, C.~Theiler, T.~Golfinopoulos, J.L. Terry, E.~Marmar, C.~Wüthrich, C.K. Tsui, H.~de~Oliveira, B.P. Duval, D.~Galassi, D.S. Oliveira, D.~Mancini, and the TCV~Team.
\newblock {Cross-field and parallel dynamics of SOL filaments in TCV}.
\newblock {\em Nucl. Fusion}, \href{https://dx.doi.org/10.1088/1741-4326/ac7ed7}{\textbf{62} 096014}, (2022).

\end{thebibliography}
